\documentclass[11pt]{article}
\usepackage[utf8]{inputenc}
\usepackage[T1]{fontenc}
\usepackage{amsmath,amssymb,amsthm}
\usepackage{booktabs}
\usepackage[margin=2.6cm]{geometry}
\usepackage{tikz}
\usetikzlibrary{arrows.meta,backgrounds}
\usepackage{xcolor}

\usepackage{hyperref}

\newtheorem{theorem}{Theorem}
\newtheorem{lemma}[theorem]{Lemma}
\newtheorem{corollary}[theorem]{Corollary}
\newtheorem{observation}[theorem]{Observation}
\theoremstyle{definition}
\newtheorem{definition}[theorem]{Definition}

\theoremstyle{remark}

\newtheorem{problem}{Open problem}

\newcommand{\pos}{\mathrm{pos}}
\newcommand{\LBb}{\mathrm{LB}_{\mathrm{bi}}}
\newcommand{\LBu}{\mathrm{LB}_{\mathrm{uni}}}
\newcommand{\G}{\mathcal{G}}
\newcommand{\cB}{\beta}
\newcommand{\no}[1]{}

\title{Tight Asymptotic Bounds on the Size of Order Graphs}
\author{Ruben Becker$^1$, Adri\'an G\'omez-Brand\'on$^2$, Gonzalo Navarro$^3$, Nicola Prezza$^1$ \\
\ \\ 
\small $^1~$ Ca' Foscari University of Venice, Italy \\
\small $^2~$ Universidade da Coru\~na \& IMFD, Spain \\
\small $^3~$ Universidad de Chile, DCC \& IMFD, Chile \\
.}
\date{}

\begin{document}
\maketitle

\begin{abstract}
Order graphs were introduced by Arroyuelo et al.\ (TODS 2024) as a generalization of the ring index, to
support worst-case optimal joins on relations of arity $d$ while storing few columns: the
index stores one column per edge, so its size is proportional to the number of edges of an order graph.
It was known that the minimum number of edges lies between $\Omega(2^d\sqrt d)$ and $O(2^d d)$, and the lower bound had been
claimed to be $42$ for $d=6$. We first exhibit an optimal order graph with only $40$ edges for $d=6$, and show
that the proof of that lower bound holds only for unions of cycles, whereas order graphs can be smaller than the
smallest set of cycles.
We give a corrected statement and proof that holds for
every order graph: there must be at least $\max_m\binom dm\lceil m/2\rceil$ edges. This bound is $40$ (hence tight) for $d=6$.
We then settle the asymptotic question: the minimum size is $\Theta(2^d\sqrt d)$.
This is obtained via an explicit construction, based on symmetric chain
decompositions of the Boolean lattice, with at most $(d-1)\binom{d}{\lfloor d/2\rfloor}+2^d$ edges.
The upper bound is within a factor $4+O(d^{-1/2})$ of the lower bound, and both numerically and asymptotically better than trie switching.
Finally, we cast the problem as a set cover with connectivity constraints, which yields an exact ILP
formulation and a greedy heuristic. With them we find optimal order graphs for $d\le 6$ in the bidirectional and for $d \le 7$ in the unidirectional model, showing that our lower bounds are tight there, and improve the best known order graphs for $d=7$ and $8$.
\end{abstract}

\section{Introduction}

\paragraph{Worst-case optimal joins and their space problem.}
A multijoin algorithm is \emph{worst-case optimal} (wco) if it runs in time proportional to the AGM bound
$Q^*$~\cite{AGM13}, the largest possible output of the query on databases of the given size. Classic multijoin algorithms that join two tables at a time are not wco, as they can produce intermediate results that are asymptotically larger than $Q^*$ \cite{AGM13}. Since the first wco
algorithm~\cite{NPRR18}, Leapfrog TrieJoin (LTJ)~\cite{Vel14} has become a reference wco algorithm because
of its simplicity. Instead of joining tables pairwise, LTJ ``binds'' one join attribute at a time: at each step, every joined relation that contains the next
attribute $x$ must be able to search the values of $x$ compatible with the attributes already bound.
With classical prefix-based indexes (tries, B-trees), this requires the bound attributes to form a prefix of an
indexed order, followed by $x$. Supporting every binding order thus requires up to $d!$ orders for a
relation of arity $d$, or the equivalent of $2^{d-1}$ tries with the \emph{trie switching} technique on meta-tries~\cite[Lemma~7.8]{AGHNRRStods24}. Since the tries have $d$ levels, they store in the worst case $d!(dN)$ values for a table of $N$ tuples, or $2^{d-1}(dN)$ with trie switching. This is much more than the $dN$ values needed to store the table in plain form.

\paragraph{The ring.}
The ring~\cite{AHNRRSsigmod21,AGHNRRStods24} avoids this redundancy. It stores each attribute as a separate
column, sorted by a stable re-sort of a ``previous'' one, which allows the rows of any column to
be mapped to those of the next efficiently without pointers. For $d \le 3$, a single
ring with $d$ columns supports LTJ in wco time over every binding order, using essentially the same $dN$ values of storage required anyway to store the table in plain form. The key property is that, regarded cyclically and in both directions, a sequence of the $d \le 3$ attributes contains every possible binding sequence: any subset of already bound attributes can be extended to contain the new attribute $x$ to bind, either backwards or forwards. The ring exploits this fact by always maintaining a range of some column that represents the current node of any of the $d!$ tries along the LTJ algorithm.

\paragraph{From rings to order graphs.}
For $d>3$ a single ring no longer suffices; for example with attributes $A,B,C,D$, the circular string $ABCD$ does not contain the substring $AC$ (nor $CA$); we need two rings like $A,B,C,D$ and $A,C,B,D$ to do the trick. The natural question is: how many rings are needed with $d$ attributes? It was proved that $O(2^d)$ rings, thereby using $O(2^d(dN))$ space, were sufficient to support LTJ~\cite{AGHNRRStods24}, which puts the ring in the same asymptotic family of trie switching.

Arroyuelo et al.~\cite{AGHNRRStods24} then generalized sets of rings to \emph{order graphs}: directed graphs whose nodes are attribute orders,
where an edge labeled by attribute $a$ goes from an order $\pi$ to the order obtained by stably re-sorting $\pi$ by $a$ (that
is, by moving $a$ to the front). For that edge, the index stores the column $a$ of the table, in the order that results from lexicographically sorting the table rows with the attributes reordered as $\pi$ (several edges may be labeled $a$, so the index will store several copies of column $a$, with orders given by different $\pi$s). To support LTJ in wco time, an order graph must be
\emph{complete}: for every set $S$ of bound attributes and every next attribute $x$, the graph must contain a path
labeled by $S\cup\{x\}$ with $x$ at one end
\cite[Thm.~7.4]{AGHNRRStods24}. Since the index stores a column ($N$ values) per graph edge, the central combinatorial question
is:
\begin{center}
\emph{What is the minimum number of edges of a complete order graph of dimension $d$?}
\end{center}
A set of rings is an order graph formed by various length-$d$ cycles, but complete order graphs can be strictly smaller: for $d=5$, a single cycle of length $20$ is complete,
whereas $5$ rings ($25$ edges) are needed if only rings of length $d$ are used.

\paragraph{The lower bound.}
The minimum size of a complete order graph was shown to lie between $\Omega(2^d\sqrt d)$ \cite[Thm.~7.9]{AGHNRRStods24} and $O(2^d d)$ (from the $O(2^d)$ rings). Exhaustive combinatorial exploration, mostly restricted to cycles and sets of rings, found that the minimal complete order graphs for $d\le 6$ matched the numerical
lower bound \cite[Thm.~7.9]{AGHNRRStods24}, including a cyclic graph of size $42$ for $d=6$. These numbers are much smaller than the sizes of trie switching, which backed the conjecture that the actual size is closer to $2^d\sqrt{d}$ than to the $2^{d-1} d$ of trie switching, and thus order graphs are indeed superior. But the conjecture remained unproved.

\paragraph{Our contributions}
\begin{enumerate}
\item \textbf{An optimal order graph for $d=6$ and a flaw in previous work.}
We present a complete order graph for $d=6$ with only $40$
edges. It is found with Gurobi on an exact ILP formulation of the optimization problem given in Section~\ref{sec:greedy}, and it is not a union of cycles (see Figure~\ref{fig:g40}); it is a single ``hairy'' cycle (i.e., with trees sprouting from its nodes). This result contradicts the published lower bound \cite[Thm.~7.9]{AGHNRRStods24}. We show that the
proof holds for unions of cycles, but not for hairy cycles, where several paths can share their final
position. Our graph is also the first known case where a complete order graph is strictly smaller than every set of
cycles, as the bound $42$ for $d=6$ does hold for those. This shows, in a sense, that the combinatorial structure of the problem is much more complicated than previously thought. 

\item \textbf{A correct lower bound} (Theorem~\ref{thm:lb}). Any complete order graph has at least $\max_m \binom dm\lceil m/2\rceil$ edges. The key
  is to charge each requirement to the \emph{last} edge of its ``witness path'', which determines the whole label
  sequence regardless of branching. This coincides with the previous bounds for $d \le 8$~\cite{AGHNRRStods24} except for $d=6$, where it gives exactly $40$. Hence our graph of Figure~\ref{fig:g40} is minimal. We also give
  a proof valid for arbitrary graphs of the unidirectional bound $d\binom{d-1}{\lfloor(d-1)/2\rfloor}$ (where $x$ can only occur after $S$, useful for restricted classes of rings \cite{AGHNRRStods24}). Both lower bounds are still $\Omega(2^d \sqrt d)$.

\item \textbf{A matching upper bound} (Theorem~\ref{thm:ub}). An explicit construction, computable in time $O(2^d d)$,
  with at most $(d-1)\binom d{\lfloor d/2\rfloor}+2^d = O(2^d \sqrt d)$ edges, which is complete even with the unidirectional restriction. This is obtained by means of symmetric chain decompositions \cite{GK76}.
Consequently, the minimum size of a complete order graph in dimension $d$ is $\Theta(2^d\sqrt d)$ in both the bidirectional and the unidirectional case, which closes the question in the asymptotic sense (Corollary~\ref{cor:asym}) and establishes the superiority over trie switching. The ratio
  between our upper and lower bounds is at most $4+O(d^{-1/2})$ in the bidirectional model and at most
  $2+O(d^{-1/2})$ in the unidirectional one.

\item \textbf{Small order graphs in practice} (Section~\ref{sec:greedy}). We cast the problem as a set cover
  with connectivity constraints on the De Bruijn graph, which yields an exact ILP formulation
  and a greedy heuristic. Reducing the ILP to the ``distinct tails'' of the strings makes it solvable for
  $d\le 6$ within a minute, which shows that both lower bounds are tight for $d\le 6$; in particular, the minimum
  unidirectional size for $d=6$ is $60$. Solving the reduced ILP with a commercial solver also reaches the
  unidirectional lower bound for $d=7$, so the minimum there is $140$. For $d=7$ and $d=8$, the ILP and the greedy
  heuristic improve the best known order graphs in both models. We list the best order graphs we found in
  Appendix~\ref{app:graphs}.
\end{enumerate}

\paragraph{Organization.}
Section~\ref{sec:prelim} fixes notation. Section~\ref{sec:lb} proves the lower bounds and explains the flaw in the
original argument. Section~\ref{sec:ub} gives the upper-bound construction, and
Section~\ref{sec:asym} derives the asymptotics and compares with known values.
Section~\ref{sec:greedy} gives an exact ILP formulation and introduces a greedy heuristic that yields much better upper bounds for small $d$.
Section~\ref{sec:open}
discusses remaining gaps and concludes.

\section{Basic Concepts}\label{sec:prelim}

Let the attribute set be $[d]=\{1,\dots,d\}$. An \emph{order} is a permutation $\pi=\pi(1)\pi(2)\cdots\pi(d)$ of $[d]$.
We write $\pi[i..j]$ for the subsequence $\pi(i)\cdots\pi(j)$, also used as a set when convenient, and
$\pos_\pi(a)=\pi^{-1}(a)$ for the position of $a$ in $\pi$. For $a\in[d]$, $a\cdot\pi$ is the order obtained by moving $a$ to
the front of $\pi$, keeping the relative order of the other attributes.

\begin{definition}[{\cite[Def.~7.1]{AGHNRRStods24}}]
An \emph{order graph} of dimension $d$ is a labeled directed graph $\G=(V,E)$ where $V$ is a set of orders and
an edge from $\pi$ to $\pi'$ with label $a$ may exist only if $\pi'=a\cdot\pi$.
\end{definition}

Note that, in an order graph, the source and the label determine the target; we write edges as $e=(\pi,a)$, always with
$a\neq\pi(1)$ (self-loops are useless). The \emph{size} of $\G$ is $|E|$.

\begin{definition}\label{def:requirement}
A {\em requirement} is a pair $(S,x)$, with $S\subsetneq[d]$ and 
$x\in[d]\setminus S$. An order graph $\G$ {\em serves} $(S,x)$ if it has 
a directed path with $|S|+1$ edges whose labels are exactly $S\cup\{x\}$, and where the edge labeled $x$ is the first 
or the last one. In the {\em unidirectional model}, the edge labeled $x$ must
be the last one in the path.
The requirement has {\em level} $m=|S|+1$ and \emph{set} $U=S\cup\{x\}$, so $|U|=m$. 
\end{definition}

Note that there are exactly $m\binom dm$ requirements
of level $m$, namely $m$ for each $m$-subset $U$ of $[d]$. To support LTJ, an
order graph must serve them all.

\begin{definition}[{\cite[Def.~7.2]{AGHNRRStods24}}]\label{def:complete}
$\G$ is \emph{complete} if it serves every possible requirement $(S,x)$.
\end{definition}

\begin{observation}[implicit in {\cite[Lemma~7.6]{AGHNRRStods24}}]\label{obs:prefix}
If a path with distinct labels $b_1,\dots,b_k$ ends at node $\nu$, then $\nu[1..k]=b_kb_{k-1}\cdots b_1$.
\end{observation}
\begin{proof}
By induction on $k$. The last edge moves $b_k$ to the front and shifts $b_{k-1},\dots,b_1$, which by induction
occupied the first $k-1$ positions and differ from $b_k$, one position to the right.
\end{proof}

\section{Lower Bounds}\label{sec:lb}

We first note that there is exactly one label sequence per edge and length.

\begin{lemma}\label{lem:one}
Let $e=(\nu,a)$ and $m\ge 1$. If a path with $m$ edges and distinct labels ends with $e$, then its label sequence
is exactly $\nu(m-1)\cdots\nu(1)\,a$, and $\pos_\nu(a)\ge m$.
\end{lemma}
\begin{proof}
The first $m-1$ edges form a path with distinct labels ending at $\nu$, so by Observation~\ref{obs:prefix} their
labels are $\nu(m-1),\dots,\nu(1)$. Since $a$ differs from all of them, $a\notin\nu[1..m-1]$, that is,
$\pos_\nu(a)\ge m$. Thus, the $m$-sequence that arrives at the target node $a\cdot\nu$ of $e$ is $\nu(m-1)\cdots\nu(1)\,a$.
\end{proof}

This implies that every edge $e=(\nu,a)$ and every $m\le\pos_\nu(a)$ together determine a unique set
\[
U_m(e)\;=\;\nu[1..m-1]\cup\{a\},
\]
and therefore a path with $m$ edges ending with $e$ can serve only two requirements $(S,x)$ of level $m$, both with set
$U_m(e)$:
\begin{itemize}
\item $(\nu[1..m-1],\,a)$, with $x=a$ last, and
\item $(\nu[1..m-2]\cup\{a\},\,\nu(m-1))$, with $x=\nu(m-1)$ first.
\end{itemize}
For $m=1$ both coincide with $(\emptyset,a)$.

\begin{theorem}[correcting {\cite[Thm.~7.9]{AGHNRRStods24}}]\label{thm:lb}
Let $\G=(V,E)$ be a complete order graph of dimension $d$, and let
$E_{\ge m}=\{(\nu,a)\in E:\pos_\nu(a)\ge m\}$. Then, for every $m\in[d]$,
\[
|E_{\ge m}|\;\ge\;\binom dm\left\lceil\frac m2\right\rceil.
\]
In particular, $|E|\ge\LBb(d):=\max_{1\le m\le d}\binom dm\lceil m/2\rceil$.
\end{theorem}
\begin{proof}
Fix $m$. For each requirement of level $m$, choose a {\em witness path} as in Definition~\ref{def:requirement} and charge
the requirement to the last edge of the path, which belongs to $E_{\ge m}$ by Lemma~\ref{lem:one}. By the same
lemma, an edge $e$ is charged at most two requirements, both with set $U_m(e)$. Therefore, the $m$ requirements
with a given set $U$ are charged to at least $\lceil m/2\rceil$ distinct edges $e$, all with $U_m(e)=U$. Since
$U_m(e)$ is determined by $e$, edges charged for different sets are distinct. Summing over the $\binom dm$ sets
$U$ yields the bound.
\end{proof}

\begin{theorem}[unidirectional case]\label{thm:lbuni}
If $\G$ is complete in the unidirectional model, then for every $m$, $|E_{\ge m}|\ge m\binom dm=d\binom{d-1}{m-1}$. Hence
$|E|\ge\LBu(d):=d\binom{d-1}{\lfloor(d-1)/2\rfloor}$.
\end{theorem}
\begin{proof}
As before, but now an edge is charged at most one requirement of each level.
\end{proof}

Note that the proofs do not depend on the shape of $\G$; in particular they need neither the reduction to in-degree~1 
\cite[Lemma~7.7]{AGHNRRStods24} nor any assumption on cycles.

\subsection{What went wrong in the original proof}

By the symmetry $\binom dl=\binom d{d-l}$, each inner term $\binom dl\lceil(d-l)/2\rceil$ of the original bound equals the term $\binom dm\lceil m/2\rceil$ of Theorem~\ref{thm:lb} with $m=d-l$, so the only numerical difference is the final rounding per attribute.
The original argument, however, fails in two steps.
\begin{enumerate}
\item It counts per $l$-set $S$, assuming that each occurrence of $S$ as a path serves at most two attributes
  $x$, one at each end. With branching this fails: if $\nu[1..l]=S$ and $\nu$ has outgoing edges $(\nu,x)$ for
  various $x\notin S$, a single occurrence of $S$ serves all those requirements $(S,x)$. Theorem~\ref{thm:lb} counts
  instead per set $U=S\cup\{x\}$ and charges it to the \emph{last} edge, which determines the whole sequence
  (Lemma~\ref{lem:one}) regardless of branching.
\item It rounds per attribute, as $d\cdot\lceil\binom dl\lceil(d-l)/2\rceil/d\rceil$. This rounding relies on the
  per-$S$ counting of the first step, and it is exactly what turns $40$ into $42$ for $d=6$.
\end{enumerate}
The value of the unidirectional bound~\cite{AGHNRRStods24} was correct; Theorem~\ref{thm:lbuni} provides a
proof valid for arbitrary order graphs. For unions of cycles, however, the original argument is valid. There, the only paths of $l+1$ edges containing a window of $l$ consecutive edges extend it by the edge just before or just after it, so each window with distinct labels serves at most two requirements $(S,x)$ with $|S|=l$. Every $l$-set $S$ must therefore appear in at least $\lceil(d-l)/2\rceil$ windows. Each edge labeled $a$ belongs to at most $l$ windows, so there are at least $\lceil\binom{d-1}{l-1}\lceil(d-l)/2\rceil/l\rceil=\lceil\binom dl\lceil(d-l)/2\rceil/d\rceil$ edges labeled $a$, for each $a$. For $d=6$ and $l=3$ this gives $6\cdot\lceil 20/3\rceil=42$.

\begin{corollary}\label{cor:d6}
The minimum size of a complete order graph of dimension $6$ is exactly $40$.
\end{corollary}
\begin{proof}
$\LBb(6)=\max\{6,15,40,30,18,3\}=40$, attained at $m=3$, and the graph of Figure~\ref{fig:g40} is complete with
$40$ edges, as can be checked directly (we also verified it mechanically, see Appendix~\ref{app:check}).
\end{proof}

Independently, the ILP of Section~\ref{sec:greedy} was solved to optimality for $d=6$, which by
Lemma~\ref{lem:ilp} certifies again that $40$ is the minimum.

For all other $d\le 8$ the corrected bound has the same value as in Arroyuelo et al.~\cite{AGHNRRStods24}; see
Table~\ref{tab:values}. In particular, the exhaustive searches done in that paper remain conclusive for $d\le 5$. The next difference with previous work occurs at $d=10$ (760 versus our 756).

\begin{figure}[p]
\centering
\begin{tikzpicture}[x=1cm,y=1cm,>=Stealth,
  nd/.style={draw,rounded corners=1.5pt,inner sep=1.4pt,font=\scriptsize\ttfamily,fill=white},
  cy/.style={nd,fill=black!12},
  el/.style={font=\scriptsize,inner sep=1pt,fill=white}]
\node[nd] (n0) at (0.00,-5.52) {621534};
\node[nd] (n1) at (1.55,-7.36) {543216};
\node[nd] (n2) at (1.55,-6.44) {432156};
\node[nd] (n3) at (3.10,-11.96) {562314};
\node[nd] (n4) at (3.10,-11.04) {623514};
\node[nd] (n5) at (3.10,-10.12) {235164};
\node[nd] (n6) at (3.10,-9.20) {351624};
\node[nd] (n7) at (4.65,-9.20) {451632};
\node[nd] (n8) at (3.88,-8.28) {516324};
\node[nd] (n9) at (3.88,-7.36) {163254};
\node[nd] (n10) at (6.20,-9.20) {124635};
\node[nd] (n11) at (7.75,-11.04) {635241};
\node[nd] (n12) at (7.75,-10.12) {352461};
\node[nd] (n13) at (7.75,-9.20) {524631};
\node[nd] (n14) at (6.98,-8.28) {246315};
\node[nd] (n15) at (9.30,-12.88) {254136};
\node[nd] (n16) at (9.30,-11.96) {541362};
\node[nd] (n17) at (9.30,-11.04) {413562};
\node[nd] (n18) at (10.85,-13.80) {342615};
\node[nd] (n19) at (10.85,-12.88) {426135};
\node[nd] (n20) at (10.85,-11.96) {261354};
\node[nd] (n21) at (12.40,-11.96) {461352};
\node[nd] (n22) at (11.62,-11.04) {613542};
\node[nd] (n23) at (10.46,-10.12) {135462};
\node[nd] (n24) at (10.46,-9.20) {354621};
\node[nd] (n25) at (10.46,-8.28) {546321};
\node[nd] (n26) at (8.72,-7.36) {463215};
\node[nd] (n27) at (6.30,-6.44) {632154};
\node[nd] (n28) at (3.92,-5.52) {321564};
\node[nd] (n29) at (1.96,-4.60) {215634};
\node[nd] (n30) at (1.96,-3.68) {156342};
\node[nd] (n31) at (1.96,-2.76) {563412};
\node[nd] (n32) at (1.96,-1.84) {634125};
\node[nd] (n33) at (1.96,-0.92) {341256};
\node[cy] (n34) at (1.52,0.00) {412563};
\node[cy] (n35) at (0.79,1.40) {641253};
\node[cy] (n36) at (2.70,2.26) {564123};
\node[cy] (n37) at (4.60,1.40) {256413};
\node[cy] (n38) at (3.87,0.00) {125643};
\node[nd] (n39) at (5.47,-0.90) {312564};
\draw[->,very thick] (n34) -- node[el] {6} (n35);
\draw[->,very thick] (n35) -- node[el] {5} (n36);
\draw[->,very thick] (n36) -- node[el] {2} (n37);
\draw[->,very thick] (n37) -- node[el] {1} (n38);
\draw[->,very thick] (n38) -- node[el] {4} (n34);
\draw[->] (n34) -- node[el] {3} (n33);
\draw[->] (n33) -- node[el] {6} (n32);
\draw[->] (n32) -- node[el] {5} (n31);
\draw[->] (n31) -- node[el] {1} (n30);
\draw[->] (n30) -- node[el] {2} (n29);
\draw[->] (n38) -- node[el] {3} (n39);
\draw[->] (n29) -- node[el] {6} (n0);
\draw[->] (n29) -- node[el] {3} (n28);
\draw[->] (n28) -- node[el] {4} (n2);
\draw[->] (n2) -- node[el] {5} (n1);
\draw[->] (n28) -- node[el] {6} (n27);
\draw[->] (n27) -- node[el] {1} (n9);
\draw[->] (n27) -- node[el] {4} (n26);
\draw[->] (n9) -- node[el] {5} (n8);
\draw[->] (n8) -- node[el] {3} (n6);
\draw[->] (n8) -- node[el] {4} (n7);
\draw[->] (n6) -- node[el] {2} (n5);
\draw[->] (n5) -- node[el] {6} (n4);
\draw[->] (n4) -- node[el] {5} (n3);
\draw[->] (n26) -- node[el] {2} (n14);
\draw[->] (n26) -- node[el] {5} (n25);
\draw[->] (n14) -- node[el] {1} (n10);
\draw[->] (n14) -- node[el] {5} (n13);
\draw[->] (n13) -- node[el] {3} (n12);
\draw[->] (n12) -- node[el] {6} (n11);
\draw[->] (n25) -- node[el] {3} (n24);
\draw[->] (n24) -- node[el] {1} (n23);
\draw[->] (n23) -- node[el] {4} (n17);
\draw[->] (n17) -- node[el] {5} (n16);
\draw[->] (n16) -- node[el] {2} (n15);
\draw[->] (n23) -- node[el] {6} (n22);
\draw[->] (n22) -- node[el] {2} (n20);
\draw[->] (n20) -- node[el] {4} (n19);
\draw[->] (n19) -- node[el] {3} (n18);
\draw[->] (n22) -- node[el] {4} (n21);
\end{tikzpicture}
\caption{An optimal complete order graph for $d=6$ with $40$ edges, generated 
with Gurobi from an ILP formulation of the optimization problem ($d=7$ turned
out to be too large to solve in the same way).
Each node shows its order and each edge its label. The cycle (gray nodes,
thick edges) has labels $6,5,2,1,4$; attribute $3$ never occurs on it, so it stays last in every cycle order.
Every other order follows from its parent by moving the edge label to the front. Every node has in-degree~$1$,
so the graph is a single ``hairy'' cycle in the sense of Arroyuelo et al.~\cite[Lemma~7.7]{AGHNRRStods24}.}
\label{fig:g40}
\end{figure}

\section{An Upper Bound of $O(2^d\sqrt d)$}\label{sec:ub}

We use a \emph{symmetric chain decomposition (SCD)} of the Boolean lattice $2^{[d]}$~\cite{dBTK51}.
The parenthesis-matching
construction of Greene and Kleitman~\cite{GK76} produces it in $O(2^d d)$ time. 

\begin{definition}
An SCD of $2^{[d]}$ is a partition of the $2^d$ subsets of $[d]$
into $M=\binom d{\lfloor d/2\rfloor}$ chains $C_1,\dots,C_M$, where $C_j$ contains exactly one set of
each size in $[\cB_j,\,d-\cB_j]$, $\cB_j$ being the size of its smallest set, and each other set of $C_j$ contains the
previous one.
\end{definition}

Note that exactly one chain, the one containing $\emptyset$ and $[d]$, has $\cB_j=0$; in general, there are
$\binom db-\binom d{b-1}$ chains with $\cB_j=b$, for $0\le b\le\lfloor d/2\rfloor$, since every chain with $\cB_j\le b$ meets level $b$ exactly once.
Let $\ell_j=d-2\cB_j+1$ be the
number of sets in $C_j$; then it follows that $\sum_j\ell_j=2^d$.

We extend each chain $C_j$ to an order $\pi_j$: first the elements of the smallest set of $C_j$ (in any order), then the
element added by each subsequent set, and finally the remaining elements. Thus every $S\in C_j$ satisfies
$S=\pi_j[1..|S|]$. Let $k_j^-=\max(1,\cB_j)$ and $k_j^+=\min(d-1,\,d-\cB_j)$.

\paragraph{The construction.} For each $j$, let $\pi=\pi_j$ and $h=k_j^+$, and add the following edges:
\begin{enumerate}
\item \textbf{History.} A path $H_j$ with $h$ edges, starting at $\nu_0=\pi[h+1..d]\;\pi[1..h]$, with labels
  $\pi(h),\pi(h-1),\dots,\pi(1)$. By induction, the node reached after $t$ edges is
  \[
  \nu_t=\pi[h-t+1..h]\;\pi[h+1..d]\;\pi[1..h-t],
  \]
  so $\nu_h=\pi$. No edge is a self-loop, because its label $\pi(h-t)$ is the last element of $\nu_t$.
\item \textbf{Hairs.} The edges $(\pi,\pi(p))$ for $p=k_j^-+1,\dots,d$.
\end{enumerate}
The order graph $\G_d$ is the union of all these edges, where repeated edges are counted once.

\begin{theorem}\label{thm:ub}
$\G_d$ is complete (even in the unidirectional model), can be built in $O(2^d d)$ time, and
\[
|E(\G_d)|\;\le\;\sum_{j=1}^M 2(d-\cB_j)\;=\;(d-1)\binom d{\lfloor d/2\rfloor}+2^d.
\]
\end{theorem}
\begin{proof}
\emph{Completeness.} For $S=\emptyset$ we need an edge labeled $x$, for every $x$. The chain $C_j$ containing $\emptyset$
has $\cB_j=0$, and at its order $\pi_j$ there are hairs labeled $\pi_j(2),\dots,\pi_j(d)$, while the last edge of
its history is labeled $\pi_j(1)$.

Let now $S\neq\emptyset$ with $1 \le |S|=k\le d-1$, and let $C_j$ be the unique chain containing $S$. Then
$\cB_j\le k\le d-\cB_j$, so $k_j^-\le k\le k_j^+$, and $S=\pi_j[1..k]$. The last $k$ edges of $H_j$ have the
distinct labels $\pi_j(k),\dots,\pi_j(1)$ and end at $\pi_j$. Since $x\notin S$, we have
$\pos_{\pi_j}(x)>k\ge k_j^-$, so the hair $(\pi_j,x)$ exists. Together they form a path with $k+1$ edges, labels
$S\cup\{x\}$, and $x$ last.

\emph{Size.} Chain $j$ contributes $k_j^+\le d-\cB_j$ history edges and $d-k_j^-\le d-\cB_j$ hairs. Since
$2(d-\cB_j)=d+\ell_j-1$, summing over the $M$ chains gives $dM+2^d-M$.

\emph{Time.} The SCD and the orders $\pi_j$ are obtained in $O(d)$ time per set of the lattice, and each chain
contributes $O(d)$ edges.
\end{proof}

\paragraph{Example.}
Let $d=4$. Figure~\ref{fig:scd4} shows the Greene--Kleitman SCD of $2^{[4]}$, both as a table and on the
Hasse diagram of the lattice, together with the resulting orders $\pi_j$.

\begin{figure}[t]
\centering
\definecolor{ch1}{HTML}{E69F00}
\definecolor{ch2}{HTML}{56B4E9}
\definecolor{ch3}{HTML}{009E73}
\definecolor{ch4}{HTML}{D55E00}
\definecolor{ch5}{HTML}{0072B2}
\definecolor{ch6}{HTML}{CC79A7}
\begin{minipage}[c]{0.46\textwidth}
\centering\small
\begin{tabular}{@{}c@{\hspace{0.5em}}l@{\hspace{0.8em}}c@{\hspace{0.8em}}c@{}}
\toprule
 & chain $C_j$ & $\pi_j$ & $[k_j^-,k_j^+]$\\
\midrule
\tikz[baseline=-0.6ex]\draw[ch1,line width=2.2pt] (0,0) -- (0.45,0); & $\emptyset\subset1\subset12\subset123\subset1234$ & $1234$ & $[1,3]$\\
\tikz[baseline=-0.6ex]\draw[ch2,line width=2.2pt] (0,0) -- (0.45,0); & $2\subset23\subset234$ & $2341$ & $[1,3]$\\
\tikz[baseline=-0.6ex]\draw[ch3,line width=2.2pt] (0,0) -- (0.45,0); & $3\subset13\subset134$ & $3142$ & $[1,3]$\\
\tikz[baseline=-0.6ex]\draw[ch4,line width=2.2pt] (0,0) -- (0.45,0); & $4\subset14\subset124$ & $4123$ & $[1,3]$\\
\tikz[baseline=-0.6ex]\draw[ch5,line width=2.2pt] (0,0) -- (0.45,0); & $24$ & $2413$ & $[2,2]$\\
\tikz[baseline=-0.6ex]\draw[ch6,line width=2.2pt] (0,0) -- (0.45,0); & $34$ & $3412$ & $[2,2]$\\
\bottomrule
\end{tabular}
\end{minipage}\hfill
\begin{minipage}[c]{0.52\textwidth}
\centering
\begin{tikzpicture}[x=0.97cm,y=1.05cm,
  st/.style={draw=#1,line width=1.2pt,fill=white,rounded corners=2pt,inner sep=1.5pt,
             minimum width=0.62cm,minimum height=0.42cm,font=\footnotesize},
  cov/.style={black!22,line width=0.5pt},
  chn/.style={line width=2.2pt}]
\node[st=ch1] (e)    at (0,0) {$\emptyset$};
\node[st=ch1] (1234) at (0,4) {$1234$};
\node[st=ch2] (2) at (-2.4,1) {$2$};
\node[st=ch3] (3) at (-0.8,1) {$3$};
\node[st=ch1] (1) at ( 0.8,1) {$1$};
\node[st=ch4] (4) at ( 2.4,1) {$4$};
\node[st=ch5] (24) at (-3.75,2) {$24$};
\node[st=ch2] (23) at (-2.25,2) {$23$};
\node[st=ch3] (13) at (-0.75,2) {$13$};
\node[st=ch1] (12) at ( 0.75,2) {$12$};
\node[st=ch4] (14) at ( 2.25,2) {$14$};
\node[st=ch6] (34) at ( 3.75,2) {$34$};
\node[st=ch2] (234) at (-2.4,3) {$234$};
\node[st=ch3] (134) at (-0.8,3) {$134$};
\node[st=ch1] (123) at ( 0.8,3) {$123$};
\node[st=ch4] (124) at ( 2.4,3) {$124$};
\begin{scope}[on background layer]
\foreach \a/\b in {e/1,e/2,e/3,e/4,
  1/12,1/13,1/14,2/12,2/23,2/24,3/13,3/23,3/34,4/14,4/24,4/34,
  12/123,12/124,13/123,13/134,14/124,14/134,23/123,23/234,24/124,24/234,34/134,34/234,
  123/1234,124/1234,134/1234,234/1234}
  \draw[cov] (\a) -- (\b);
\draw[chn,ch1] (e) -- (1) -- (12) -- (123) -- (1234);
\draw[chn,ch2] (2) -- (23) -- (234);
\draw[chn,ch3] (3) -- (13) -- (134);
\draw[chn,ch4] (4) -- (14) -- (124);
\end{scope}
\end{tikzpicture}
\end{minipage}
\caption{The Greene--Kleitman SCD of $2^{[4]}$ used in our example, with sets written without braces. Left: the
six chains $C_j$, their orders $\pi_j$, and the ranges $[k_j^-,k_j^+]$. Right: the Hasse diagram of $2^{[4]}$, where
each chain is a path drawn in its color (the chains $\{2,4\}$ and $\{3,4\}$ are single nodes) and gray lines are
the remaining inclusions. Each order $\pi_j$ lists the elements in the order in which its chain adds them.}
\label{fig:scd4}
\end{figure}

The histories of $1234$, $2341$, $4123$, and $3412$ all lie on the ring $1234\to4123\to3412\to2341\to1234$,
whose consecutive orders are rotations of each other. The history of $3142$ closes a second ring, because its
hair labeled $2$ leads back to the start $2314$ of its history. The bound of Theorem~\ref{thm:ub} is $34$ and the
construction produces $32$ edges (the bound overcounts 2 edges in the chain that contains $\emptyset$ and $[d]$), but repeated edges reduce $\G_4$ to $21$ edges (Figure~\ref{fig:d4}).

Consider $S=\{1,3\}$ and $x=4$. The chain containing $S$ is $C_3$, with $\pi_3=3142$, so $S=\pi_3[1..2]$. The
witness consists of the last two edges of the history of $3142$, namely $4231\xrightarrow{1}1423\xrightarrow{3}3142$,
followed by the hair $3142\xrightarrow{4}4312$. Its labels are
$1,3,4$, with $x=4$ last.

\begin{figure}[t]
\centering
\begin{tikzpicture}[x=1cm,y=1cm,>=Stealth,
  nd/.style={draw,rounded corners=1.5pt,inner sep=1.6pt,font=\footnotesize\ttfamily,fill=white},
  pj/.style={nd,fill=black!15},
  el/.style={font=\scriptsize,inner sep=1pt,fill=white}]
\node[pj] (a1) at (0,2) {1234};
\node[pj] (a2) at (2.6,2) {4123};
\node[pj] (a3) at (2.6,0) {3412};
\node[pj] (a4) at (0,0) {2341};
\draw[->,very thick] (a1) -- node[el]{4} (a2);
\draw[->,very thick] (a2) -- node[el]{3} (a3);
\draw[->,very thick] (a3) -- node[el]{2} (a4);
\draw[->,very thick] (a4) -- node[el]{1} (a1);
\node[nd] (b1) at (5.4,2) {1423};
\node[pj] (b2) at (8.0,2) {3142};
\node[nd] (b3) at (8.0,0) {2314};
\node[nd] (b4) at (5.4,0) {4231};
\draw[->] (b1) -- node[el]{3} (b2);
\draw[->] (b2) -- node[el]{2} (b3);
\draw[->] (b3) -- node[el]{4} (b4);
\draw[->] (b4) -- node[el]{1} (b1);
\node[pj] (c1) at (4.0,3.7) {2413};
\node[nd] (c2) at (6.7,3.7) {4132};
\node[nd] (c3) at (9.3,3.7) {1324};
\draw[->] (c3) -- node[el]{4} (c2);
\draw[->] (c2) -- node[el]{2} (c1);
\draw[->] (a2) -- node[el]{1} (b1);
\draw[->] (a2) -- node[el]{2} (c1);
\draw[->] (a4) .. controls (1.2,-1.4) and (4.2,-1.4) .. node[el,pos=0.5]{4} (b4);
\node[nd] (l1) at (-1.9,2.6) {2134};
\node[nd] (l2) at (-1.9,1.4) {3124};
\draw[->] (a1) -- node[el]{2} (l1);
\draw[->] (a1) -- node[el]{3} (l2);
\node[nd] (l3) at (-1.9,-0.6) {3241};
\draw[->] (a4) -- node[el]{3} (l3);
\node[nd] (l4) at (4.0,0.9) {1342};
\draw[->] (a3) -- node[el]{1} (l4);
\node[nd] (l5) at (9.9,2.6) {1342};
\node[nd] (l6) at (9.9,1.4) {4312};
\draw[->] (b2) -- node[el]{1} (l5);
\draw[->] (b2) -- node[el]{4} (l6);
\node[nd] (l7) at (2.2,4.6) {1243};
\node[nd] (l8) at (4.9,5.0) {3241};
\draw[->] (c1) -- node[el]{1} (l7);
\draw[->] (c1) -- node[el]{3} (l8);
\end{tikzpicture}
\caption{The graph $\G_4$ of our example, with $21$ edges. Gray nodes are the chain orders $\pi_j$;
thick edges form the ring shared by four histories. Leaves reached by two edges ($1342$ and $3241$) are drawn
twice, which does not change the edge count. The four edges $1324\to4132\to2413$ and $2314\to4231\to1423$ are
redundant: the paths $1234\xrightarrow{4}4123\xrightarrow{2}2413$ and $1234\xrightarrow{4}4123\xrightarrow{1}1423$
already provide those histories.}
\label{fig:d4}
\end{figure}

\section{Asymptotics and Exact Values}\label{sec:asym}

We use the central binomial estimate $\binom n{\lfloor n/2\rfloor}=\bigl(\sqrt{2/\pi}+o(1)\bigr)\,2^n/\sqrt n$ from Stirling's approximation, and write
$B=\binom{d-1}{\lfloor(d-1)/2\rfloor}$. By Pascal's triangle rule, $\binom d{\lfloor d/2\rfloor}\le 2B$.

\begin{lemma}\label{lem:asym}
For every $d\ge 2$,
    \[
	    \tfrac d2\,B\;\le\;\LBb(d)\;\le\;\tfrac d2\,B+\tfrac12\binom d{\lfloor d/2\rfloor},\qquad
	    \LBu(d)=d\,B,\qquad
	    |E(\G_d)|\;\le\;2(d-1)B+2^d .
	    \]
	    Consequently,
	    \begin{align*}
	    \LBb(d)&=\Bigl(\frac{1}{2\sqrt{2\pi}}+o(1)\Bigr)2^d\sqrt d,\qquad
	    \LBu(d)=\Bigl(\frac{1}{\sqrt{2\pi}}+o(1)\Bigr)2^d\sqrt d,\\
		    |E(\G_d)|&\le\Bigl(\sqrt{\frac2\pi}+o(1)\Bigr)2^d\sqrt d .
		    \end{align*}
		    \end{lemma}
		    \begin{proof}
		    Since $\binom dm\frac m2=\frac d2\binom{d-1}{m-1}$, we have
		    $\frac d2\binom{d-1}{m-1}\le\binom dm\lceil m/2\rceil\le\frac d2\binom{d-1}{m-1}+\frac12\binom dm$, and the
		    maximum of $\binom{d-1}{m-1}$ over $m$ is $B$. The value of $\LBu$ is its definition. The bound on $|E(\G_d)|$
		    follows from Theorem~\ref{thm:ub} and $\binom d{\lfloor d/2\rfloor}\le 2B$. The asymptotic forms follow from
		    $B=\bigl(\sqrt{2/\pi}+o(1)\bigr)2^{d-1}/\sqrt d$ and $2^d=o(2^d\sqrt d)$.
		    \end{proof}

		    \begin{corollary}\label{cor:asym}
		    The minimum size of a complete order graph of dimension $d$ is $\Theta(2^d\sqrt d)$, both in the bidirectional
		    and in the unidirectional model. More precisely, for every $d\ge2$,
		    \[
			    \frac{|E(\G_d)|}{\LBb(d)}\;\le\;4+O(d^{-1/2}),\qquad\frac{|E(\G_d)|}{\LBu(d)}\;\le\;2+O(d^{-1/2}).
			    \]
			    This improves the previous upper bound of $O(2^d d)$ \cite{AGHNRRStods24} by a factor $\Theta(\sqrt d)$.
			    \end{corollary}
			    \begin{proof}
			    By Lemma~\ref{lem:asym}, $|E(\G_d)|/\LBb(d)\le\bigl(2(d-1)B+2^d\bigr)/(\frac d2B)<4+2^{d+1}/(dB)$, and
			    $2^{d+1}/(dB)=O(d^{-1/2})$ because $B=\Theta(2^d/\sqrt d)$. The unidirectional case is identical, with
			    $\LBu(d)=dB$.
			    \end{proof}

			    Table~\ref{tab:values} compares values for small $d$. The construction does not compete with the solutions found by computational
			    search~\cite{AGHNRRStods24} for $d\le 8$; its interest is asymptotic. Still, it is smaller than the trie-switching
			    bound $2^{d-1}d$~\cite[Lemma~7.8]{AGHNRRStods24} for every $d\ge 3$. In the next section we introduce a heuristic that matches or improves the best known values in most cases.

				    \begin{table}[t]
\centering
\small
\setlength{\tabcolsep}{4.5pt}
\begin{tabular}{rrrrrrrrrrr}
\toprule
 & \multicolumn{4}{c}{Bidirectional} & \multicolumn{3}{c}{Unidirectional} & \multicolumn{3}{c}{Formal constructions}\\
\cmidrule(lr){2-5}\cmidrule(lr){6-8}\cmidrule(lr){9-11}
$d$ & \cite[Thm.~7.9]{AGHNRRStods24} & $\LBb$ & previous & new & $\LBu$ & previous & new & $|E(\G_d)|$ & Thm.~\ref{thm:ub} & $2^{d-1}d$\\
\midrule
3 & 3   & 3   & 3   & 3            & 6   & 6   & 6             & 6   & 14  & 12\\
4 & 8   & 8   & 8   & 8            & 12  & 15  & \textbf{12}   & 21  & 34  & 32\\
5 & 20  & 20  & 20  & 20           & 30  & 33  & \textbf{30}   & 52  & 72  & 80\\
6 & 42  & \textbf{40}  & 42  & \textbf{40}  & 60  & 66  & \textbf{60}   & 132 & 164 & 192\\
7 & 70  & 70  & 83  & \textbf{77}  & 140 & 162 & \textbf{140}  & 300 & 338 & 448\\
8 & 168 & 168 & 200 & \textbf{187} & 280 & 360 & \textbf{297}  & 697 & 746 & 1024\\
{9} & {378} & {378} & {--} & \textbf{406} & {630} & {--} & \textbf{656} & {1443} & {1520} & {2304}\\
\bottomrule
\end{tabular}
\caption{Lower bounds, known sizes, and formal constructions. ``Previous'' is taken from Table~6
of~\cite{AGHNRRStods24}. ``New'' is the best size we found (Appendix~\ref{app:graphs}): with the ILP of
Section~\ref{sec:greedy}, where the bidirectional values for $d\le 6$ and the unidirectional ones for $d \le 7$ are optimal; the others were found by a large neighborhood search, except for $d=9$, which were found by the greedy heuristic. Bold values improve the previous ones.
$|E(\G_d)|$ is the actual size of our formal construction of Section~\ref{sec:ub} and the next column its bound.
The last column is the trie-switching size~\cite[Lemma~7.8]{AGHNRRStods24}.}
\label{tab:values}
\end{table}

\section{A Set-Cover Formulation and a Greedy Heuristic}\label{sec:greedy}

The construction of Section~\ref{sec:ub} is within a constant factor of optimal, but for small $d$ it is far from
the best known order graphs (Table~\ref{tab:values}). In this section we reformulate the problem as a set cover with
connectivity constraints on the De Bruijn graph. This yields an exact ILP formulation, which we used to find the
graph of Figure~\ref{fig:g40} and, after a reduction, to find optimal order graphs for $d\le 6$. The formulation also yields a
greedy heuristic that finds much smaller complete order graphs than $\G_d$ for $d\le 8$, within minutes.

\subsection{The De Bruijn formulation}

There is always a minimum complete order graph with all its in-degrees equal to  $1$~\cite[Lemma~7.7]{AGHNRRStods24}. In such a
graph, every node has a unique, infinite, backward walk, and we can identify the node with the string of the last
$d$ labels read along that walk. These strings are the nodes of the De Bruijn graph.

\paragraph{De Bruijn graph.}
Let $D=(V_D,E_D)$ be the De Bruijn graph with $V_D=[d]^d$ and an edge from $v$ to $w$, labeled $w[d]$, whenever
$v[2..d]=w[1..d-1]$. For $u\in V_D$, $N^-(u)$ denotes its set of predecessors.

\paragraph{Coverage.}
A string $u\in V_D$ \emph{covers} a requirement $(S,x)$ of level $m=|S|+1$ if the last $m$ symbols of $u$ are a
witness for $(S,x)$, that is, they are distinct and form $S\cup\{x\}$ with $x$ first or last. Let $\Pi(S)$ be the set
of permutations of $S$; then the strings covering $(S,x)$ are
\[
  \mathrm{Cov}(S,x)=\{\, u\in V_D :
  u[d-m+1..d]\in\{\sigma x,\ x\sigma : \sigma\in\Pi(S)\} \,\}.
\]
In the unidirectional model, only the strings $\sigma x$ are allowed. Note that a string covers at most two
requirements of each level, so at most $2d$ requirements in total.

\paragraph{ILP formulation.}
We use one binary variable $z_u$ per string $u\in V_D$, which is $1$ if $u$ is \emph{selected}:
\begin{align}
  \min\quad & \sum_{u\in V_D} z_u \\
  \text{s.t.}\quad & \sum_{u\in \mathrm{Cov}(S,x)} z_u \;\ge\; 1
      && \forall (S,x), \label{eq:cover}\\
  & \sum_{v\in N^-(u)} z_v \;\ge\; z_u
      && \forall u\in V_D, \label{eq:indeg}\\
  & z_u\in\{0,1\} && \forall u\in V_D.
\end{align}
Constraint~\eqref{eq:cover} asks every requirement to be covered by a selected string, and
constraint~\eqref{eq:indeg} asks every selected string to have a selected predecessor.

\begin{lemma}\label{lem:ilp}
The optimum of the ILP equals the minimum size of a complete order graph of dimension $d$. Moreover, any set of
strings satisfying Eqs.~\eqref{eq:cover} and~\eqref{eq:indeg} can be decoded, in linear time, into a complete order graph
with at most as many edges as strings.
\end{lemma}
\begin{proof}
Let $z$ be a feasible solution. Keep exactly one edge of $D$ entering each selected node, from a selected
predecessor (it exists by Eq.~\eqref{eq:indeg}). The result has $\sum_u z_u$ edges and every node has in-degree $1$,
so from every node $u$ there is an infinite backward walk, whose last $d$ labels are $u$ by the definition of $D$.
Assign to $u$ the order formed by the distinct labels of its backward walk, sorted by their most recent occurrence,
followed by the attributes that never occur in the walk, in increasing order. If $(v,u)$ is an edge with label $c$,
the walk of $u$ is that of $v$ preceded by $c$, so the order of $u$ is $c\cdot\pi_v$, where $\pi_v$ is the order of $v$:
the edges are valid order-graph edges (nodes receiving the same order are merged, and edges with $c=\pi_v(1)$
become self-loops and are dropped). If $u\in\mathrm{Cov}(S,x)$, the last $m$ edges of the walk into $u$ form a path
with distinct labels $S\cup\{x\}$ and $x$ at one end, so $(S,x)$ is served. Hence the result is a complete order
graph with at most $\sum_u z_u$ edges.

Conversely~\cite[Lemma~7.7]{AGHNRRStods24}, there is a minimum complete order graph in which every node has
in-degree exactly $1$. Select, for each node, the string of the last $d$ labels of its (unique and infinite) backward
walk. This selects at most $|V|=|E|$ nodes of $D$; the selected string of the predecessor of a node is a predecessor
in $D$ of the selected string of the node, so Eq.~\eqref{eq:indeg} holds; and a path serving $(S,x)$ and ending at a
node $u$ is a suffix of the backward walk of $u$, so Eq.~\eqref{eq:cover} holds.
\end{proof}

The ILP has $d^d$ variables, which allowed Gurobi to solve it for $d=6$ ($46{,}656$ variables), yielding the
graph of Figure~\ref{fig:g40}, but not for $d=7$ ($823{,}543$ variables): its LP bound stayed at
$\LBb(7)=70$, while the large neighborhood search built into Gurobi, warm-started from the best solution found
by the ILP solver, found a complete order graph with $77$ edges, which is still the best known for $d=7$
(Appendix~\ref{app:graphs}). A large neighborhood search repeatedly frees part of the current solution and
rebuilds it optimally, by solving a smaller ILP; it often improves solutions, but provides no lower bounds.

\subsection{Reducing the ILP to distinct tails}

Many strings of $D$ are equivalent for the ILP. The \emph{distinct tail} $\tau(u)$ of $u\in V_D$ is its longest
suffix formed by distinct symbols; for example, $\tau(3141526)=41526$. Let $T$ be the set of nonempty strings of
distinct symbols of $[d]$, and for $t\in T$ and $c\in[d]$ let $\delta(t,c)\in T$ be the distinct tail of $t\cdot c$,
that is, $t\cdot c$ if $c\notin t$, and the part of $t$ that follows $c$, followed by $c$, otherwise. We say that $t\in T$
covers $(S,x)$ if its last $m=|S|+1$ symbols are a witness for $(S,x)$, as for the strings of $V_D$.

\begin{lemma}\label{lem:tails}
Consider the ILP with one binary variable $y_t$ per $t\in T$, the covering constraints of Eq.~\eqref{eq:cover} on $T$, and
the constraints $\sum_{t'} y_{t'}\ge y_t$ for every $t\in T$, where $t'\ne t$ ranges over the strings with
$\delta(t',c)=t$ and $c$ is the last symbol of $t$. Its optimum equals that of the ILP on $V_D$.
\end{lemma}
\begin{proof}
A string $u\in V_D$ covers the same requirements as $\tau(u)$, because witnesses consist of distinct symbols. If $D$
has an edge from $v$ to $u$ with label $c$, then $\tau(u)=\delta(\tau(v),c)$: if $c\notin\tau(v)$, then $|\tau(v)|<d$,
so $\tau(v)$ is a suffix of $v[2..d]$ and $\tau(u)=\tau(v)\cdot c$ (a longer distinct suffix of $u$ would yield one of $v$
longer than $\tau(v)$); otherwise, the occurrence of $c$ in $\tau(v)$ bounds $\tau(u)$. Hence, mapping a feasible
solution on $V_D$ to the set of tails of its selected strings gives a feasible solution on $T$ that is not larger.

Conversely, given a feasible solution on $T$, choose a selected predecessor for each selected $t$, and let $u(t)$
be the string of the last $d$ symbols read along the resulting infinite backward walk from $t$, where each node
contributes its last symbol. The walk spells $t$ in its last $|t|$ symbols, and the symbol read before them belongs to
$t$: when $t=\delta(t',c)$ with $c\in t'$ it is $c$ itself, and when $c\notin t'$ the same holds for $t'\subset t$ by
induction (the length cannot decrease indefinitely). Thus $\tau(u(t))=t$, so the strings $u(t)$ are distinct, cover the
same requirements, and $u(t')$ is a predecessor of $u(t)$ in $D$. This is a feasible solution on $V_D$ of the same
size.
\end{proof}

The new ILP has $|T|=\sum_{k=1}^{d}d!/(d-k)!=d!\sum_{j=0}^{d-1}1/j! < e\cdot d!$ variables, that is, $1{,}956$, $13{,}699$, and
$109{,}600$ for $d=6,7,8$. We strengthen it with two families of valid constraints. First, for each $m$-set $U$, the
selected tails whose last $m$ symbols form $U$ must number at least $\lceil m/2\rceil$ (at least $m$ in the
unidirectional model), by the argument of Theorem~\ref{thm:lb}. Second, some selected tail must have length $d$ to
cover the requirements of level $d$, and by renaming the attributes we may assume that $y_{12\cdots d}=1$.

With these additions, the free solver HiGHS solves the ILP for every $d\le 6$, in both models, in less than a
minute. This confirms that $40$ is the minimum for $d=6$ in the bidirectional model, and shows that the
unidirectional lower bound is also attained for $d=6$. For $d=7$, HiGHS did not close the gap within
$40$ minutes. 
The branch-and-bound solver of Gurobi, instead, found a unidirectional solution with $140=\LBu(7)$ edges, which
is then optimal. For $d=8$, it did not reach the lower bounds either, but its large neighborhood search, warm-started
from the best solutions found by branch and bound, reached $187$ and $297$ edges in the bidirectional and
unidirectional models after some hours.
In the bidirectional model for $d=7$, the gap between $70$ and $77$ remains open.
We obtain the following (see Table~\ref{tab:values}).

\begin{corollary}\label{cor:uni}
In the unidirectional model, the minimum size of a complete order graph is $\LBu(d)$ for $d\le 7$, namely $6$,
$12$, $30$, $60$, and $140$ for $d=3$ to $7$.
\end{corollary}

\subsection{A greedy heuristic}

The ILP gives optimal or near-optimal order graphs, but its cost grows quickly with $d$: even after the
reduction to distinct tails, it has $\lfloor e\cdot d!\rfloor-1$ variables, and for $d=8$ it needed hours of large
neighborhood search to improve its solutions. We now give a much cheaper greedy heuristic, which does not compete
with the ILP when the latter can be solved, but runs within minutes for $d=8$ and reaches larger values of $d$.

Without the constraint of Eq.~\eqref{eq:indeg}, the ILP is a plain set cover instance, and the classical greedy algorithm would
apply. The heuristic below adapts it so that the constraint holds at all times. It maintains a set $X$
of selected strings, initially the $d$ rotations of $12\cdots d$, which form a cycle of $D$ (a ring), so every
string of $X$ has a predecessor in $X$. Then:

\begin{enumerate}
\item For each string $s\notin X$, let $o<d$ be the largest value such that some string $t \in X$ ends with
  $s[1..o]$. Adding $s$ to $X$ then requires adding the at most $d-o$ strings along the path of $D$ that extends $t$ with $s[o+1..d]$, the last of which is $s$; we call them the \emph{path} of $s$.
\item Add to $X$ the path of the string $s$ that maximizes the number of newly covered requirements (by all the
  strings of its path) divided by the length of its path, breaking ties at random. Repeat until all the requirements
  are covered.
\item Remove from $X$, in random order, every string whose removal keeps all the requirements covered and retains that every
  remaining string has a predecessor in $X$.
\item Decode $X$ into an order graph as in the proof of Lemma~\ref{lem:ilp}, and verify its completeness.
\end{enumerate}

Step~1 is the key difference with the plain greedy set cover, after which each selected string would have to be
connected separately by a path of up to $d$ new strings: here a new string is connected through its longest overlap
with the strings already selected, and the strings added on the way count towards the coverage.

\paragraph{Example.}
Let $d=4$, so there are $32$ requirements. In the bidirectional model, the initial ring $X=\{1234,2341,3412,4123\}$
covers $24$ of them. In the first iteration, the string $s=1231$ overlaps in $o=3$ symbols with $t=4123\in X$, so its path
is just $s$, which covers $3$ new requirements: $(\{1\},3)$ and $(\{3\},1)$, with witness $31$, and $(\{1,3\},2)$, with
witness $231$ and $x$ first. Three more iterations of this kind add $2342$, $2314$, and $3142$, and the resulting $8$
strings decode into an optimal complete order graph with $8$ edges. In the unidirectional model, the ring covers only
$16$ requirements and the heuristic needs $7$ iterations. In the second one, the string $s=1243$ overlaps in only
$o=2$ symbols with $t=3412\in X$, so its path is $4124,1243$; the intermediate string $4124$ covers $(\{2\},4)$ and
$(\{1,2\},4)$, and $s$ covers $(\{4\},3)$ and $(\{2,4\},3)$, for a ratio of $4/2=2$. The final $12$ strings decode into
an optimal unidirectional order graph (Corollary~\ref{cor:uni}).

\paragraph{Results.}
Each iteration evaluates all the $d^d$ strings, so the heuristic takes $O(d^{d+1})$ time per selected string; with a
straightforward C++ implementation, a run takes a few seconds for $d=7$ and a few minutes for $d=8$. We report the
best results over $8$--$24$ random seeds. The heuristic attains the optimum for $d\le 5$ in both models. In the
bidirectional model, it finds $41$, $88$, and $189$ edges for $d=6$, $7$, and $8$, against $40$, $77$, and $187$
obtained with the ILP; in the unidirectional model, it finds $63$, $141$, and $302$ edges, against $60$, $140$, and
$297$. Except for the bidirectional case $d=7$, the heuristic is thus within $5\%$ of the ILP solutions, at a tiny
fraction of their cost. Its solutions are also much smaller than those known before~\cite{AGHNRRStods24}, namely $42$,
$83$, and $200$ in the bidirectional model and $66$, $162$, and $360$ in the unidirectional one, which were obtained
by searching (exhaustively or heuristically) only unions of rings and single cycles.

For $d=9$, where the reduced ILP has almost $10^6$ variables, the heuristic finds complete order graphs with
406 and 656 edges in the bidirectional and unidirectional models, respectively,
against lower bounds of $\LBb(9)=378$ and $\LBu(9)=630$.

\section{Conclusions and Open Problems}\label{sec:open}

We have shown that the minimum number of columns of an order index supporting worst-case optimal joins on a
relation of arity $d$ is $\Theta(2^d\sqrt d)$. This closes the gap between the $\Omega(2^d\sqrt d)$ and $O(2^d d)$
bounds known so far and shows that order graphs can use asymptotically less space than trie switching. On the way, we found a complete order graph for $d=6$ with $40$ edges, smaller than any union
of cycles, which showed that the previous lower bound argument applied only to unions of cycles. We then gave a
lower bound valid for every order graph, which proves that this graph is optimal. Our upper bound is an explicit
construction based on symmetric chain decompositions of the Boolean lattice. It uses only the unidirectional model, 
so it also settles this case asymptotically. 
The remaining gaps are constant factors: asymptotically, the upper and
lower bounds differ by a factor of $4$ in the bidirectional model and $2$ in the unidirectional one. In broad
terms, an order index stores $\Theta(2^d/\sqrt d)$ times more data than the relation itself, against $d!$ for classical indexes over all orders and $\Theta(2^d)$ with trie switching on meta-tries. 
We also gave an exact ILP formulation which, reduced to distinct tails, finds optimal order graphs for $d\le 6$
within a minute, and for $d=7$ in the unidirectional model with a commercial solver. It also yields a greedy
heuristic that is within $1\%$ of the unidirectional lower bound for $d=7$. Together, they improve the best known
order graphs for $d=7$ and $8$.
These results considerably extend our understanding of order graphs and
confirm that they can make worst-case optimal joins affordable for moderate arities. We finish with some open problems.

\paragraph{Sharing histories (a factor of 2).}
Half of the cost of $\G_d$ is spent on histories, paid separately for each $\pi_j$. Our example shows how
they can often be shared, and the near-optimal unidirectional graphs found by the heuristic of
Section~\ref{sec:greedy} show that much better sharing is possible; the question is whether it can be achieved by an
explicit construction. In general, suppose that $\pi_j=a\cdot\pi_i$ with $\pos_{\pi_i}(a)\ge k_j^+$ and
$k_i^+\ge k_j^+-1$. Then the edge $(\pi_i,a)$, which is a hair of $\pi_i$ whenever $\pos_{\pi_i}(a)>k_i^-$, preceded
by the history of $\pi_i$, is already a valid history for $\pi_j$, and $H_j$ can be dropped. 
The histories then form a forest rooted at the chains whose histories cannot be shared. After removing redundant
edges as in~\cite[Lemma~7.6]{AGHNRRStods24}, each tree can be closed into a hairy cycle of the same size, as in
the proof of~\cite[Lemma~7.7]{AGHNRRStods24}.

\begin{problem}
Is there an SCD of $2^{[d]}$, and extensions $\pi_j$ of its chains, forming an arborescence in the sense above
except for $o(M)$ roots? Then the hairs, $\sum_j(d-k_j^-)\le\frac d2M+2^{d-1}$, would dominate the size, yielding
complete order graphs with $(1+o(1))\LBu(d)$ edges. This would settle the unidirectional case save for lower-order terms.
\end{problem}

\paragraph{Using extensions in both directions (the other factor of 2).}
To reach $\LBb$, almost every edge $e$ must serve both requirements of $U_m(e)$, at every central level
$m=d/2\pm O(\sqrt d)$. In $\G_d$, hairs serve only requirements in one direction.

\begin{problem}
Construct complete order graphs with $(1+o(1))\LBb(d)$ edges, or improve the lower bound by showing that the
history constraints prevent a constant fraction of the edges from serving both ends.
\end{problem}

\paragraph{Tightness of the unidirectional bound.}
The ILP of Section~\ref{sec:greedy} shows that $\LBu(d)$ is attained for $d\le 7$, and the best graph we
found for $d=8$ is $6\%$ above it ($297$ versus $280$).

\begin{problem}
Is $\LBu(d)=d\binom{d-1}{\lfloor(d-1)/2\rfloor}$ the minimum size of a complete order graph in the unidirectional
model, for every $d$? Or at least, is the minimum $(1+o(1))\LBu(d)$?
\end{problem}

\paragraph{Rings.}
The bound of~\cite[Lemma~6.8]{AGHNRRStods24} gives $\Omega(2^d/\sqrt d)$ rings of length $d$, while the best known
upper bound is $O(2^d)$ rings.

\begin{problem}
Is there a family of $O(2^d/\sqrt d)$ cyclic orders of $[d]$ such that every pair $(U,x)$ with $x\in U$ appears, in
some of them, as an arc with set $U$ and $x$ at one end? If so, sets of rings also reach $\Theta(2^d\sqrt d)$ edges.
\end{problem}

\paragraph{Paths with repeated labels.}
Definition~\ref{def:requirement} asks for paths with $|S|+1$ edges, hence without repeated labels. The order index
could also use longer paths that repeat labels, by restricting again by an attribute that is already bound, at
some cost in query time~\cite{AGHNRRStods24}. Exhaustive searches found no case where this yields smaller
complete order graphs~\cite{AGHNRRStods24}.
Observation~\ref{obs:prefix} generalizes to this case: if a path ends at $\nu$, then $\nu$ starts with the distinct
labels of the path, sorted by their last occurrence, most recent first. Hence, if $x$ occurs only as the last
label of the path, the last edge $(\nu,x)$ determines the set $\nu[1..m-1]\cup\{x\}$, and
Theorem~\ref{thm:lbuni} still holds. When $x$ is the first label but occurs again later in the path, however, $x$
need not be at position $m-1$ of $\nu$. Then a single edge might serve more than two requirements with the same set,
and the proof of Theorem~\ref{thm:lb} no longer applies.

\begin{problem}
Does Theorem~\ref{thm:lb} hold for bidirectional order graphs when paths may repeat labels? More generally, can
repeated labels ever reduce the minimum size of a complete order graph?
\end{problem}

\bibliographystyle{plain}
\bibliography{refs}

\appendix
\section{Mechanical Verification}\label{app:check}

We implemented a checker for Definition~\ref{def:complete}, which computes, for each node, the longest path
reaching it whose labels form a reversed prefix of its order, and then the requirements served by each edge
according to Lemma~\ref{lem:one}. 
With it we verified that
\begin{itemize}
\item the graph of Figure~\ref{fig:g40} is complete with $40$ edges;
\item $\G_d$ is complete for $d=3,\dots,9$, with the sizes of Table~\ref{tab:values} ($1443$ edges for $d=9$);
\item the $21$-edge graph of Figure~\ref{fig:d4} is complete;
\item the graphs found by the heuristic of Section~\ref{sec:greedy} (Table~\ref{tab:values}) are complete in
their respective models;
\item the graphs listed in Appendix~\ref{app:graphs} are complete in their respective models, and have the sizes
given there.
\end{itemize}

\section{Best Order Graphs Found}\label{app:graphs}

We list the smallest complete order graphs we found, which give the ``new'' columns of Table~\ref{tab:values}.
All of them were verified mechanically (Appendix~\ref{app:check}). Each line describes a weakly connected component:
a start order $\pi_0$, followed by a bracketed list of edges. A digit $a$ is an edge labeled $a$ from the current node
to $a\cdot\pi$, where $\pi$ is the order of the current node, and the brackets following it list the edges leaving
that target; a digit with a star is an edge whose target already appeared (such as the edge closing a cycle). For
example, from $\pi_0=123456$, the string \texttt{3[6[5]2]4} describes the edges
$123456\xrightarrow{3}312456$, $312456\xrightarrow{6}631245$, $631245\xrightarrow{5}563124$,
$312456\xrightarrow{2}231456$, and $123456\xrightarrow{4}412356$. The following Python function lists the edges of a
component, as triples (source, label, target) of strings:
\begin{verbatim}
def edges(start, s):
    out, stack, cur = [], [], start
    for c in s:
        if c.isdigit():
            last = c + cur.replace(c, "")
            out.append((cur, c, last))
        elif c == "[":
            stack.append(cur)
            cur = last
        elif c == "]":
            cur = stack.pop()
    return out
\end{verbatim}

\paragraph{Bidirectional model.}
\noindent$d=3$, $3$ edges:\\[2pt]
{\small\raggedright\texttt{132:\ 2\allowbreak{}[\allowbreak{}3\allowbreak{}[\allowbreak{}1\allowbreak{}$^*$\allowbreak{}]\allowbreak{}]\allowbreak{}}\par}\smallskip
\noindent$d=4$, $8$ edges:\\[2pt]
{\small\raggedright\texttt{3214:\ 1\allowbreak{}[\allowbreak{}4\allowbreak{}[\allowbreak{}2\allowbreak{}]\allowbreak{}]\allowbreak{}4\allowbreak{}[\allowbreak{}1\allowbreak{}[\allowbreak{}2\allowbreak{}[\allowbreak{}3\allowbreak{}$^*$\allowbreak{}]\allowbreak{}]\allowbreak{}2\allowbreak{}]\allowbreak{}}\par}\smallskip
\noindent$d=5$, $20$ edges:\\[2pt]
{\small\raggedright{\texttt{15432:\ 2\allowbreak{}[\allowbreak{}3\allowbreak{}[\allowbreak{}1\allowbreak{}[\allowbreak{}4\allowbreak{}]\allowbreak{}4\allowbreak{}[\allowbreak{}1\allowbreak{}[\allowbreak{}2\allowbreak{}5\allowbreak{}]\allowbreak{}2\allowbreak{}5\allowbreak{}[\allowbreak{}1\allowbreak{}$^*$\allowbreak{}2\allowbreak{}]\allowbreak{}]\allowbreak{}5\allowbreak{}[\allowbreak{}4\allowbreak{}]\allowbreak{}]\allowbreak{}]\allowbreak{}3\allowbreak{}[\allowbreak{}5\allowbreak{}[\allowbreak{}2\allowbreak{}[\allowbreak{}1\allowbreak{}4\allowbreak{}[\allowbreak{}1\allowbreak{}]\allowbreak{}]\allowbreak{}]\allowbreak{}]\allowbreak{}}}\par}\smallskip
\noindent$d=6$, $40$ edges (that of Figure~\ref{fig:g40}):\\[2pt]
{\small\raggedright{\texttt{412563:\ 3\allowbreak{}[\allowbreak{}6\allowbreak{}[\allowbreak{}5\allowbreak{}[\allowbreak{}1\allowbreak{}[\allowbreak{}2\allowbreak{}[\allowbreak{}3\allowbreak{}[\allowbreak{}4\allowbreak{}[\allowbreak{}5\allowbreak{}]\allowbreak{}6\allowbreak{}[\allowbreak{}1\allowbreak{}[\allowbreak{}5\allowbreak{}[\allowbreak{}3\allowbreak{}[\allowbreak{}2\allowbreak{}[\allowbreak{}6\allowbreak{}[\allowbreak{}5\allowbreak{}]\allowbreak{}]\allowbreak{}]\allowbreak{}4\allowbreak{}]\allowbreak{}]\allowbreak{}4\allowbreak{}[\allowbreak{}2\allowbreak{}[\allowbreak{}1\allowbreak{}5\allowbreak{}[\allowbreak{}3\allowbreak{}[\allowbreak{}6\allowbreak{}]\allowbreak{}]\allowbreak{}]\allowbreak{}5\allowbreak{}[\allowbreak{}3\allowbreak{}[\allowbreak{}1\allowbreak{}[\allowbreak{}4\allowbreak{}[\allowbreak{}5\allowbreak{}[\allowbreak{}2\allowbreak{}]\allowbreak{}]\allowbreak{}6\allowbreak{}[\allowbreak{}2\allowbreak{}[\allowbreak{}4\allowbreak{}[\allowbreak{}3\allowbreak{}]\allowbreak{}]\allowbreak{}4\allowbreak{}]\allowbreak{}]\allowbreak{}]\allowbreak{}]\allowbreak{}]\allowbreak{}]\allowbreak{}]\allowbreak{}6\allowbreak{}]\allowbreak{}]\allowbreak{}]\allowbreak{}]\allowbreak{}]\allowbreak{}6\allowbreak{}[\allowbreak{}5\allowbreak{}[\allowbreak{}2\allowbreak{}[\allowbreak{}1\allowbreak{}[\allowbreak{}3\allowbreak{}4\allowbreak{}$^*$\allowbreak{}]\allowbreak{}]\allowbreak{}]\allowbreak{}]\allowbreak{}}}\par}\smallskip
\noindent$d=7$, $77$ edges:\\[2pt]
{\small\raggedright{\texttt{1237654:\ 4\allowbreak{}[\allowbreak{}3\allowbreak{}[\allowbreak{}6\allowbreak{}[\allowbreak{}5\allowbreak{}[\allowbreak{}2\allowbreak{}[\allowbreak{}1\allowbreak{}[\allowbreak{}7\allowbreak{}[\allowbreak{}3\allowbreak{}[\allowbreak{}5\allowbreak{}[\allowbreak{}4\allowbreak{}]\allowbreak{}]\allowbreak{}]\allowbreak{}]\allowbreak{}7\allowbreak{}]\allowbreak{}]\allowbreak{}]\allowbreak{}]\allowbreak{}]\allowbreak{}6\allowbreak{}[\allowbreak{}3\allowbreak{}[\allowbreak{}7\allowbreak{}[\allowbreak{}4\allowbreak{}[\allowbreak{}5\allowbreak{}[\allowbreak{}1\allowbreak{}[\allowbreak{}2\allowbreak{}[\allowbreak{}3\allowbreak{}[\allowbreak{}4\allowbreak{}[\allowbreak{}5\allowbreak{}[\allowbreak{}6\allowbreak{}[\allowbreak{}7\allowbreak{}[\allowbreak{}1\allowbreak{}[\allowbreak{}3\allowbreak{}[\allowbreak{}4\allowbreak{}[\allowbreak{}5\allowbreak{}]\allowbreak{}]\allowbreak{}]\allowbreak{}2\allowbreak{}[\allowbreak{}1\allowbreak{}]\allowbreak{}3\allowbreak{}[\allowbreak{}2\allowbreak{}[\allowbreak{}1\allowbreak{}$^*$\allowbreak{}]\allowbreak{}]\allowbreak{}]\allowbreak{}]\allowbreak{}]\allowbreak{}7\allowbreak{}[\allowbreak{}1\allowbreak{}[\allowbreak{}2\allowbreak{}5\allowbreak{}[\allowbreak{}6\allowbreak{}[\allowbreak{}4\allowbreak{}[\allowbreak{}2\allowbreak{}[\allowbreak{}3\allowbreak{}[\allowbreak{}1\allowbreak{}]\allowbreak{}]\allowbreak{}]\allowbreak{}]\allowbreak{}]\allowbreak{}]\allowbreak{}]\allowbreak{}]\allowbreak{}6\allowbreak{}[\allowbreak{}4\allowbreak{}[\allowbreak{}7\allowbreak{}[\allowbreak{}2\allowbreak{}[\allowbreak{}3\allowbreak{}[\allowbreak{}5\allowbreak{}[\allowbreak{}1\allowbreak{}[\allowbreak{}4\allowbreak{}[\allowbreak{}6\allowbreak{}]\allowbreak{}6\allowbreak{}[\allowbreak{}2\allowbreak{}[\allowbreak{}4\allowbreak{}[\allowbreak{}7\allowbreak{}[\allowbreak{}5\allowbreak{}]\allowbreak{}]\allowbreak{}]\allowbreak{}]\allowbreak{}]\allowbreak{}]\allowbreak{}]\allowbreak{}6\allowbreak{}[\allowbreak{}3\allowbreak{}]\allowbreak{}]\allowbreak{}]\allowbreak{}]\allowbreak{}]\allowbreak{}]\allowbreak{}]\allowbreak{}]\allowbreak{}2\allowbreak{}[\allowbreak{}3\allowbreak{}[\allowbreak{}6\allowbreak{}]\allowbreak{}6\allowbreak{}]\allowbreak{}]\allowbreak{}]\allowbreak{}]\allowbreak{}]\allowbreak{}7\allowbreak{}[\allowbreak{}4\allowbreak{}[\allowbreak{}5\allowbreak{}]\allowbreak{}5\allowbreak{}]\allowbreak{}]\allowbreak{}}}\par}\smallskip
{\small\raggedright{\texttt{5724163:\ 1\allowbreak{}[\allowbreak{}3\allowbreak{}[\allowbreak{}6\allowbreak{}]\allowbreak{}]\allowbreak{}3\allowbreak{}[\allowbreak{}6\allowbreak{}[\allowbreak{}1\allowbreak{}[\allowbreak{}4\allowbreak{}[\allowbreak{}2\allowbreak{}[\allowbreak{}5\allowbreak{}7\allowbreak{}[\allowbreak{}5\allowbreak{}$^*$\allowbreak{}]\allowbreak{}]\allowbreak{}7\allowbreak{}]\allowbreak{}]\allowbreak{}]\allowbreak{}]\allowbreak{}}}\par}\smallskip
\noindent{$d=8$, $187$ edges:}\\[2pt]
{\small\raggedright{\texttt{14367258:\ 2\allowbreak{}[\allowbreak{}7\allowbreak{}[\allowbreak{}5\allowbreak{}[\allowbreak{}4\allowbreak{}[\allowbreak{}1\allowbreak{}[\allowbreak{}6\allowbreak{}[\allowbreak{}2\allowbreak{}[\allowbreak{}8\allowbreak{}[\allowbreak{}7\allowbreak{}[\allowbreak{}3\allowbreak{}[\allowbreak{}1\allowbreak{}[\allowbreak{}6\allowbreak{}[\allowbreak{}5\allowbreak{}[\allowbreak{}2\allowbreak{}[\allowbreak{}3\allowbreak{}[\allowbreak{}1\allowbreak{}7\allowbreak{}[\allowbreak{}8\allowbreak{}]\allowbreak{}8\allowbreak{}]\allowbreak{}4\allowbreak{}[\allowbreak{}7\allowbreak{}[\allowbreak{}1\allowbreak{}3\allowbreak{}[\allowbreak{}1\allowbreak{}6\allowbreak{}[\allowbreak{}5\allowbreak{}[\allowbreak{}8\allowbreak{}]\allowbreak{}]\allowbreak{}]\allowbreak{}8\allowbreak{}[\allowbreak{}3\allowbreak{}[\allowbreak{}2\allowbreak{}6\allowbreak{}[\allowbreak{}4\allowbreak{}[\allowbreak{}1\allowbreak{}2\allowbreak{}[\allowbreak{}1\allowbreak{}]\allowbreak{}5\allowbreak{}[\allowbreak{}8\allowbreak{}[\allowbreak{}2\allowbreak{}7\allowbreak{}[\allowbreak{}2\allowbreak{}[\allowbreak{}1\allowbreak{}]\allowbreak{}]\allowbreak{}]\allowbreak{}]\allowbreak{}]\allowbreak{}]\allowbreak{}]\allowbreak{}]\allowbreak{}]\allowbreak{}]\allowbreak{}]\allowbreak{}]\allowbreak{}]\allowbreak{}]\allowbreak{}]\allowbreak{}4\allowbreak{}[\allowbreak{}5\allowbreak{}[\allowbreak{}3\allowbreak{}[\allowbreak{}1\allowbreak{}2\allowbreak{}6\allowbreak{}[\allowbreak{}2\allowbreak{}[\allowbreak{}1\allowbreak{}[\allowbreak{}8\allowbreak{}[\allowbreak{}3\allowbreak{}[\allowbreak{}4\allowbreak{}]\allowbreak{}]\allowbreak{}]\allowbreak{}]\allowbreak{}]\allowbreak{}]\allowbreak{}]\allowbreak{}]\allowbreak{}]\allowbreak{}]\allowbreak{}]\allowbreak{}8\allowbreak{}[\allowbreak{}7\allowbreak{}[\allowbreak{}3\allowbreak{}]\allowbreak{}]\allowbreak{}]\allowbreak{}8\allowbreak{}[\allowbreak{}2\allowbreak{}[\allowbreak{}7\allowbreak{}]\allowbreak{}]\allowbreak{}]\allowbreak{}]\allowbreak{}]\allowbreak{}]\allowbreak{}]\allowbreak{}5\allowbreak{}[\allowbreak{}8\allowbreak{}[\allowbreak{}6\allowbreak{}[\allowbreak{}2\allowbreak{}[\allowbreak{}3\allowbreak{}[\allowbreak{}1\allowbreak{}[\allowbreak{}7\allowbreak{}[\allowbreak{}5\allowbreak{}[\allowbreak{}2\allowbreak{}[\allowbreak{}6\allowbreak{}[\allowbreak{}8\allowbreak{}[\allowbreak{}1\allowbreak{}4\allowbreak{}[\allowbreak{}1\allowbreak{}[\allowbreak{}3\allowbreak{}7\allowbreak{}[\allowbreak{}5\allowbreak{}[\allowbreak{}3\allowbreak{}[\allowbreak{}8\allowbreak{}[\allowbreak{}2\allowbreak{}[\allowbreak{}4\allowbreak{}[\allowbreak{}1\allowbreak{}[\allowbreak{}6\allowbreak{}[\allowbreak{}7\allowbreak{}[\allowbreak{}3\allowbreak{}[\allowbreak{}2\allowbreak{}8\allowbreak{}]\allowbreak{}]\allowbreak{}]\allowbreak{}]\allowbreak{}6\allowbreak{}[\allowbreak{}7\allowbreak{}[\allowbreak{}5\allowbreak{}]\allowbreak{}]\allowbreak{}]\allowbreak{}]\allowbreak{}4\allowbreak{}[\allowbreak{}6\allowbreak{}]\allowbreak{}]\allowbreak{}]\allowbreak{}]\allowbreak{}]\allowbreak{}]\allowbreak{}7\allowbreak{}]\allowbreak{}]\allowbreak{}]\allowbreak{}]\allowbreak{}]\allowbreak{}8\allowbreak{}[\allowbreak{}4\allowbreak{}[\allowbreak{}2\allowbreak{}3\allowbreak{}[\allowbreak{}2\allowbreak{}[\allowbreak{}7\allowbreak{}[\allowbreak{}1\allowbreak{}[\allowbreak{}6\allowbreak{}[\allowbreak{}8\allowbreak{}]\allowbreak{}]\allowbreak{}]\allowbreak{}]\allowbreak{}7\allowbreak{}[\allowbreak{}5\allowbreak{}[\allowbreak{}2\allowbreak{}[\allowbreak{}1\allowbreak{}]\allowbreak{}]\allowbreak{}]\allowbreak{}]\allowbreak{}]\allowbreak{}5\allowbreak{}[\allowbreak{}1\allowbreak{}[\allowbreak{}2\allowbreak{}[\allowbreak{}6\allowbreak{}[\allowbreak{}7\allowbreak{}[\allowbreak{}5\allowbreak{}[\allowbreak{}3\allowbreak{}[\allowbreak{}1\allowbreak{}]\allowbreak{}4\allowbreak{}]\allowbreak{}]\allowbreak{}]\allowbreak{}]\allowbreak{}3\allowbreak{}[\allowbreak{}6\allowbreak{}]\allowbreak{}]\allowbreak{}4\allowbreak{}[\allowbreak{}2\allowbreak{}[\allowbreak{}6\allowbreak{}[\allowbreak{}3\allowbreak{}[\allowbreak{}1\allowbreak{}[\allowbreak{}8\allowbreak{}[\allowbreak{}7\allowbreak{}]\allowbreak{}]\allowbreak{}8\allowbreak{}[\allowbreak{}5\allowbreak{}7\allowbreak{}[\allowbreak{}5\allowbreak{}[\allowbreak{}1\allowbreak{}[\allowbreak{}2\allowbreak{}[\allowbreak{}3\allowbreak{}[\allowbreak{}8\allowbreak{}]\allowbreak{}]\allowbreak{}6\allowbreak{}[\allowbreak{}3\allowbreak{}[\allowbreak{}4\allowbreak{}]\allowbreak{}8\allowbreak{}[\allowbreak{}3\allowbreak{}[\allowbreak{}2\allowbreak{}4\allowbreak{}]\allowbreak{}]\allowbreak{}]\allowbreak{}]\allowbreak{}6\allowbreak{}]\allowbreak{}]\allowbreak{}]\allowbreak{}]\allowbreak{}]\allowbreak{}]\allowbreak{}]\allowbreak{}]\allowbreak{}]\allowbreak{}]\allowbreak{}]\allowbreak{}7\allowbreak{}]\allowbreak{}]\allowbreak{}4\allowbreak{}7\allowbreak{}[\allowbreak{}2\allowbreak{}[\allowbreak{}4\allowbreak{}[\allowbreak{}1\allowbreak{}[\allowbreak{}5\allowbreak{}]\allowbreak{}3\allowbreak{}]\allowbreak{}]\allowbreak{}4\allowbreak{}[\allowbreak{}1\allowbreak{}[\allowbreak{}3\allowbreak{}[\allowbreak{}5\allowbreak{}[\allowbreak{}2\allowbreak{}]\allowbreak{}8\allowbreak{}[\allowbreak{}5\allowbreak{}[\allowbreak{}2\allowbreak{}[\allowbreak{}1\allowbreak{}6\allowbreak{}7\allowbreak{}[\allowbreak{}6\allowbreak{}[\allowbreak{}3\allowbreak{}[\allowbreak{}4\allowbreak{}[\allowbreak{}1\allowbreak{}$^*$\allowbreak{}2\allowbreak{}]\allowbreak{}]\allowbreak{}]\allowbreak{}]\allowbreak{}]\allowbreak{}]\allowbreak{}]\allowbreak{}]\allowbreak{}5\allowbreak{}]\allowbreak{}3\allowbreak{}]\allowbreak{}]\allowbreak{}]\allowbreak{}]\allowbreak{}]\allowbreak{}}}\par}\smallskip
{\small\raggedright{\texttt{76543218:\ 1\allowbreak{}[\allowbreak{}2\allowbreak{}[\allowbreak{}4\allowbreak{}[\allowbreak{}8\allowbreak{}[\allowbreak{}5\allowbreak{}[\allowbreak{}3\allowbreak{}[\allowbreak{}6\allowbreak{}[\allowbreak{}1\allowbreak{}]\allowbreak{}]\allowbreak{}]\allowbreak{}]\allowbreak{}]\allowbreak{}]\allowbreak{}]\allowbreak{}8\allowbreak{}[\allowbreak{}1\allowbreak{}[\allowbreak{}2\allowbreak{}[\allowbreak{}3\allowbreak{}[\allowbreak{}4\allowbreak{}[\allowbreak{}5\allowbreak{}[\allowbreak{}6\allowbreak{}[\allowbreak{}1\allowbreak{}7\allowbreak{}$^*$\allowbreak{}]\allowbreak{}8\allowbreak{}[\allowbreak{}1\allowbreak{}]\allowbreak{}]\allowbreak{}]\allowbreak{}]\allowbreak{}5\allowbreak{}[\allowbreak{}4\allowbreak{}[\allowbreak{}3\allowbreak{}6\allowbreak{}]\allowbreak{}]\allowbreak{}]\allowbreak{}]\allowbreak{}]\allowbreak{}}}\par}\smallskip
\noindent{$d=9$, $406$ edges:}\\[2pt]
{\small\raggedright{\texttt{219876543:\ 3\allowbreak{}[\allowbreak{}4\allowbreak{}[\allowbreak{}5\allowbreak{}[\allowbreak{}6\allowbreak{}[\allowbreak{}1\allowbreak{}[\allowbreak{}9\allowbreak{}[\allowbreak{}2\allowbreak{}[\allowbreak{}4\allowbreak{}8\allowbreak{}[\allowbreak{}4\allowbreak{}7\allowbreak{}]\allowbreak{}]\allowbreak{}]\allowbreak{}]\allowbreak{}7\allowbreak{}[\allowbreak{}8\allowbreak{}[\allowbreak{}1\allowbreak{}[\allowbreak{}2\allowbreak{}[\allowbreak{}3\allowbreak{}]\allowbreak{}3\allowbreak{}[\allowbreak{}4\allowbreak{}[\allowbreak{}2\allowbreak{}6\allowbreak{}]\allowbreak{}]\allowbreak{}5\allowbreak{}[\allowbreak{}4\allowbreak{}[\allowbreak{}3\allowbreak{}[\allowbreak{}9\allowbreak{}[\allowbreak{}6\allowbreak{}[\allowbreak{}2\allowbreak{}[\allowbreak{}7\allowbreak{}8\allowbreak{}[\allowbreak{}4\allowbreak{}[\allowbreak{}5\allowbreak{}]\allowbreak{}]\allowbreak{}]\allowbreak{}7\allowbreak{}]\allowbreak{}]\allowbreak{}]\allowbreak{}6\allowbreak{}7\allowbreak{}9\allowbreak{}[\allowbreak{}6\allowbreak{}[\allowbreak{}8\allowbreak{}[\allowbreak{}1\allowbreak{}]\allowbreak{}]\allowbreak{}]\allowbreak{}]\allowbreak{}]\allowbreak{}9\allowbreak{}]\allowbreak{}9\allowbreak{}[\allowbreak{}1\allowbreak{}[\allowbreak{}2\allowbreak{}$^*$\allowbreak{}3\allowbreak{}[\allowbreak{}4\allowbreak{}[\allowbreak{}5\allowbreak{}6\allowbreak{}[\allowbreak{}2\allowbreak{}[\allowbreak{}5\allowbreak{}7\allowbreak{}[\allowbreak{}5\allowbreak{}[\allowbreak{}9\allowbreak{}[\allowbreak{}1\allowbreak{}]\allowbreak{}]\allowbreak{}9\allowbreak{}[\allowbreak{}8\allowbreak{}]\allowbreak{}]\allowbreak{}]\allowbreak{}7\allowbreak{}]\allowbreak{}]\allowbreak{}]\allowbreak{}4\allowbreak{}]\allowbreak{}2\allowbreak{}[\allowbreak{}3\allowbreak{}]\allowbreak{}3\allowbreak{}[\allowbreak{}1\allowbreak{}]\allowbreak{}]\allowbreak{}]\allowbreak{}9\allowbreak{}[\allowbreak{}1\allowbreak{}[\allowbreak{}3\allowbreak{}[\allowbreak{}2\allowbreak{}[\allowbreak{}6\allowbreak{}[\allowbreak{}5\allowbreak{}]\allowbreak{}]\allowbreak{}]\allowbreak{}]\allowbreak{}]\allowbreak{}]\allowbreak{}8\allowbreak{}[\allowbreak{}9\allowbreak{}[\allowbreak{}3\allowbreak{}]\allowbreak{}]\allowbreak{}9\allowbreak{}]\allowbreak{}7\allowbreak{}[\allowbreak{}8\allowbreak{}9\allowbreak{}[\allowbreak{}8\allowbreak{}[\allowbreak{}6\allowbreak{}[\allowbreak{}2\allowbreak{}]\allowbreak{}]\allowbreak{}]\allowbreak{}]\allowbreak{}8\allowbreak{}]\allowbreak{}6\allowbreak{}[\allowbreak{}8\allowbreak{}]\allowbreak{}]\allowbreak{}5\allowbreak{}[\allowbreak{}6\allowbreak{}7\allowbreak{}[\allowbreak{}8\allowbreak{}[\allowbreak{}9\allowbreak{}[\allowbreak{}1\allowbreak{}]\allowbreak{}]\allowbreak{}]\allowbreak{}]\allowbreak{}7\allowbreak{}[\allowbreak{}8\allowbreak{}[\allowbreak{}9\allowbreak{}]\allowbreak{}]\allowbreak{}]\allowbreak{}4\allowbreak{}[\allowbreak{}3\allowbreak{}5\allowbreak{}[\allowbreak{}6\allowbreak{}8\allowbreak{}]\allowbreak{}7\allowbreak{}[\allowbreak{}3\allowbreak{}[\allowbreak{}5\allowbreak{}6\allowbreak{}[\allowbreak{}2\allowbreak{}[\allowbreak{}5\allowbreak{}]\allowbreak{}8\allowbreak{}[\allowbreak{}1\allowbreak{}2\allowbreak{}5\allowbreak{}9\allowbreak{}]\allowbreak{}]\allowbreak{}8\allowbreak{}[\allowbreak{}5\allowbreak{}[\allowbreak{}1\allowbreak{}[\allowbreak{}6\allowbreak{}[\allowbreak{}4\allowbreak{}[\allowbreak{}2\allowbreak{}7\allowbreak{}[\allowbreak{}2\allowbreak{}[\allowbreak{}8\allowbreak{}]\allowbreak{}8\allowbreak{}[\allowbreak{}1\allowbreak{}]\allowbreak{}9\allowbreak{}[\allowbreak{}2\allowbreak{}[\allowbreak{}1\allowbreak{}3\allowbreak{}[\allowbreak{}5\allowbreak{}]\allowbreak{}5\allowbreak{}[\allowbreak{}3\allowbreak{}[\allowbreak{}8\allowbreak{}[\allowbreak{}6\allowbreak{}]\allowbreak{}]\allowbreak{}8\allowbreak{}[\allowbreak{}1\allowbreak{}6\allowbreak{}[\allowbreak{}1\allowbreak{}4\allowbreak{}[\allowbreak{}1\allowbreak{}[\allowbreak{}9\allowbreak{}]\allowbreak{}7\allowbreak{}]\allowbreak{}]\allowbreak{}]\allowbreak{}]\allowbreak{}]\allowbreak{}3\allowbreak{}[\allowbreak{}2\allowbreak{}[\allowbreak{}1\allowbreak{}6\allowbreak{}[\allowbreak{}5\allowbreak{}]\allowbreak{}8\allowbreak{}[\allowbreak{}1\allowbreak{}]\allowbreak{}]\allowbreak{}8\allowbreak{}[\allowbreak{}2\allowbreak{}[\allowbreak{}4\allowbreak{}5\allowbreak{}[\allowbreak{}1\allowbreak{}[\allowbreak{}4\allowbreak{}[\allowbreak{}3\allowbreak{}7\allowbreak{}]\allowbreak{}7\allowbreak{}9\allowbreak{}[\allowbreak{}4\allowbreak{}[\allowbreak{}3\allowbreak{}[\allowbreak{}7\allowbreak{}[\allowbreak{}8\allowbreak{}]\allowbreak{}]\allowbreak{}]\allowbreak{}]\allowbreak{}]\allowbreak{}4\allowbreak{}[\allowbreak{}6\allowbreak{}[\allowbreak{}7\allowbreak{}[\allowbreak{}1\allowbreak{}]\allowbreak{}]\allowbreak{}7\allowbreak{}9\allowbreak{}[\allowbreak{}6\allowbreak{}[\allowbreak{}1\allowbreak{}[\allowbreak{}7\allowbreak{}]\allowbreak{}7\allowbreak{}[\allowbreak{}8\allowbreak{}]\allowbreak{}]\allowbreak{}7\allowbreak{}[\allowbreak{}1\allowbreak{}[\allowbreak{}2\allowbreak{}[\allowbreak{}5\allowbreak{}8\allowbreak{}]\allowbreak{}3\allowbreak{}[\allowbreak{}6\allowbreak{}]\allowbreak{}8\allowbreak{}]\allowbreak{}8\allowbreak{}[\allowbreak{}2\allowbreak{}[\allowbreak{}5\allowbreak{}]\allowbreak{}]\allowbreak{}]\allowbreak{}]\allowbreak{}]\allowbreak{}]\allowbreak{}6\allowbreak{}[\allowbreak{}1\allowbreak{}4\allowbreak{}[\allowbreak{}1\allowbreak{}[\allowbreak{}3\allowbreak{}[\allowbreak{}5\allowbreak{}[\allowbreak{}2\allowbreak{}]\allowbreak{}]\allowbreak{}7\allowbreak{}[\allowbreak{}3\allowbreak{}[\allowbreak{}9\allowbreak{}]\allowbreak{}]\allowbreak{}]\allowbreak{}7\allowbreak{}]\allowbreak{}5\allowbreak{}[\allowbreak{}1\allowbreak{}7\allowbreak{}[\allowbreak{}1\allowbreak{}[\allowbreak{}3\allowbreak{}4\allowbreak{}[\allowbreak{}3\allowbreak{}[\allowbreak{}8\allowbreak{}[\allowbreak{}9\allowbreak{}]\allowbreak{}9\allowbreak{}[\allowbreak{}2\allowbreak{}[\allowbreak{}5\allowbreak{}[\allowbreak{}1\allowbreak{}6\allowbreak{}]\allowbreak{}]\allowbreak{}]\allowbreak{}]\allowbreak{}9\allowbreak{}[\allowbreak{}2\allowbreak{}[\allowbreak{}5\allowbreak{}[\allowbreak{}8\allowbreak{}]\allowbreak{}]\allowbreak{}]\allowbreak{}]\allowbreak{}9\allowbreak{}]\allowbreak{}3\allowbreak{}9\allowbreak{}[\allowbreak{}3\allowbreak{}[\allowbreak{}4\allowbreak{}8\allowbreak{}]\allowbreak{}]\allowbreak{}]\allowbreak{}]\allowbreak{}]\allowbreak{}]\allowbreak{}]\allowbreak{}]\allowbreak{}8\allowbreak{}]\allowbreak{}]\allowbreak{}]\allowbreak{}7\allowbreak{}9\allowbreak{}[\allowbreak{}4\allowbreak{}[\allowbreak{}2\allowbreak{}[\allowbreak{}3\allowbreak{}[\allowbreak{}7\allowbreak{}8\allowbreak{}[\allowbreak{}1\allowbreak{}[\allowbreak{}9\allowbreak{}[\allowbreak{}6\allowbreak{}[\allowbreak{}7\allowbreak{}[\allowbreak{}2\allowbreak{}]\allowbreak{}]\allowbreak{}]\allowbreak{}]\allowbreak{}]\allowbreak{}]\allowbreak{}7\allowbreak{}[\allowbreak{}8\allowbreak{}]\allowbreak{}8\allowbreak{}[\allowbreak{}7\allowbreak{}[\allowbreak{}1\allowbreak{}3\allowbreak{}[\allowbreak{}1\allowbreak{}[\allowbreak{}5\allowbreak{}[\allowbreak{}4\allowbreak{}]\allowbreak{}9\allowbreak{}[\allowbreak{}5\allowbreak{}]\allowbreak{}]\allowbreak{}6\allowbreak{}[\allowbreak{}9\allowbreak{}[\allowbreak{}2\allowbreak{}]\allowbreak{}]\allowbreak{}]\allowbreak{}5\allowbreak{}[\allowbreak{}1\allowbreak{}[\allowbreak{}3\allowbreak{}[\allowbreak{}6\allowbreak{}[\allowbreak{}4\allowbreak{}[\allowbreak{}8\allowbreak{}]\allowbreak{}9\allowbreak{}[\allowbreak{}2\allowbreak{}8\allowbreak{}[\allowbreak{}2\allowbreak{}[\allowbreak{}1\allowbreak{}5\allowbreak{}]\allowbreak{}4\allowbreak{}[\allowbreak{}5\allowbreak{}]\allowbreak{}]\allowbreak{}]\allowbreak{}]\allowbreak{}]\allowbreak{}9\allowbreak{}[\allowbreak{}3\allowbreak{}[\allowbreak{}6\allowbreak{}[\allowbreak{}4\allowbreak{}]\allowbreak{}]\allowbreak{}4\allowbreak{}6\allowbreak{}]\allowbreak{}]\allowbreak{}3\allowbreak{}[\allowbreak{}6\allowbreak{}[\allowbreak{}4\allowbreak{}9\allowbreak{}[\allowbreak{}1\allowbreak{}[\allowbreak{}7\allowbreak{}[\allowbreak{}4\allowbreak{}]\allowbreak{}]\allowbreak{}]\allowbreak{}]\allowbreak{}]\allowbreak{}6\allowbreak{}]\allowbreak{}]\allowbreak{}]\allowbreak{}]\allowbreak{}3\allowbreak{}[\allowbreak{}8\allowbreak{}[\allowbreak{}2\allowbreak{}[\allowbreak{}5\allowbreak{}]\allowbreak{}5\allowbreak{}]\allowbreak{}]\allowbreak{}]\allowbreak{}]\allowbreak{}]\allowbreak{}9\allowbreak{}]\allowbreak{}9\allowbreak{}[\allowbreak{}2\allowbreak{}6\allowbreak{}[\allowbreak{}1\allowbreak{}2\allowbreak{}[\allowbreak{}1\allowbreak{}[\allowbreak{}3\allowbreak{}[\allowbreak{}4\allowbreak{}[\allowbreak{}7\allowbreak{}]\allowbreak{}8\allowbreak{}[\allowbreak{}5\allowbreak{}]\allowbreak{}]\allowbreak{}4\allowbreak{}[\allowbreak{}8\allowbreak{}[\allowbreak{}3\allowbreak{}[\allowbreak{}5\allowbreak{}[\allowbreak{}1\allowbreak{}6\allowbreak{}[\allowbreak{}1\allowbreak{}[\allowbreak{}2\allowbreak{}[\allowbreak{}4\allowbreak{}[\allowbreak{}7\allowbreak{}]\allowbreak{}7\allowbreak{}]\allowbreak{}]\allowbreak{}9\allowbreak{}]\allowbreak{}]\allowbreak{}]\allowbreak{}5\allowbreak{}9\allowbreak{}[\allowbreak{}3\allowbreak{}]\allowbreak{}]\allowbreak{}]\allowbreak{}7\allowbreak{}[\allowbreak{}3\allowbreak{}[\allowbreak{}4\allowbreak{}5\allowbreak{}[\allowbreak{}4\allowbreak{}[\allowbreak{}8\allowbreak{}[\allowbreak{}6\allowbreak{}]\allowbreak{}]\allowbreak{}9\allowbreak{}[\allowbreak{}4\allowbreak{}[\allowbreak{}2\allowbreak{}[\allowbreak{}6\allowbreak{}[\allowbreak{}3\allowbreak{}]\allowbreak{}]\allowbreak{}6\allowbreak{}[\allowbreak{}8\allowbreak{}[\allowbreak{}1\allowbreak{}[\allowbreak{}2\allowbreak{}[\allowbreak{}5\allowbreak{}[\allowbreak{}7\allowbreak{}]\allowbreak{}]\allowbreak{}]\allowbreak{}2\allowbreak{}7\allowbreak{}[\allowbreak{}2\allowbreak{}[\allowbreak{}5\allowbreak{}]\allowbreak{}3\allowbreak{}]\allowbreak{}]\allowbreak{}]\allowbreak{}8\allowbreak{}[\allowbreak{}1\allowbreak{}[\allowbreak{}6\allowbreak{}[\allowbreak{}3\allowbreak{}[\allowbreak{}2\allowbreak{}[\allowbreak{}7\allowbreak{}[\allowbreak{}5\allowbreak{}[\allowbreak{}4\allowbreak{}9\allowbreak{}[\allowbreak{}1\allowbreak{}[\allowbreak{}8\allowbreak{}[\allowbreak{}3\allowbreak{}[\allowbreak{}6\allowbreak{}]\allowbreak{}6\allowbreak{}[\allowbreak{}2\allowbreak{}[\allowbreak{}7\allowbreak{}[\allowbreak{}3\allowbreak{}9\allowbreak{}[\allowbreak{}1\allowbreak{}]\allowbreak{}]\allowbreak{}]\allowbreak{}]\allowbreak{}]\allowbreak{}]\allowbreak{}3\allowbreak{}8\allowbreak{}]\allowbreak{}]\allowbreak{}9\allowbreak{}]\allowbreak{}]\allowbreak{}]\allowbreak{}7\allowbreak{}[\allowbreak{}2\allowbreak{}3\allowbreak{}[\allowbreak{}2\allowbreak{}[\allowbreak{}8\allowbreak{}[\allowbreak{}5\allowbreak{}]\allowbreak{}]\allowbreak{}4\allowbreak{}[\allowbreak{}5\allowbreak{}9\allowbreak{}]\allowbreak{}5\allowbreak{}9\allowbreak{}[\allowbreak{}5\allowbreak{}[\allowbreak{}6\allowbreak{}[\allowbreak{}4\allowbreak{}[\allowbreak{}2\allowbreak{}]\allowbreak{}]\allowbreak{}]\allowbreak{}]\allowbreak{}]\allowbreak{}]\allowbreak{}]\allowbreak{}7\allowbreak{}]\allowbreak{}2\allowbreak{}[\allowbreak{}1\allowbreak{}3\allowbreak{}[\allowbreak{}6\allowbreak{}[\allowbreak{}5\allowbreak{}[\allowbreak{}4\allowbreak{}9\allowbreak{}[\allowbreak{}1\allowbreak{}]\allowbreak{}]\allowbreak{}]\allowbreak{}7\allowbreak{}]\allowbreak{}]\allowbreak{}]\allowbreak{}]\allowbreak{}]\allowbreak{}]\allowbreak{}]\allowbreak{}4\allowbreak{}[\allowbreak{}8\allowbreak{}[\allowbreak{}2\allowbreak{}[\allowbreak{}1\allowbreak{}]\allowbreak{}3\allowbreak{}[\allowbreak{}6\allowbreak{}[\allowbreak{}5\allowbreak{}9\allowbreak{}]\allowbreak{}]\allowbreak{}5\allowbreak{}[\allowbreak{}9\allowbreak{}[\allowbreak{}1\allowbreak{}[\allowbreak{}2\allowbreak{}[\allowbreak{}7\allowbreak{}[\allowbreak{}3\allowbreak{}]\allowbreak{}]\allowbreak{}]\allowbreak{}]\allowbreak{}]\allowbreak{}9\allowbreak{}[\allowbreak{}3\allowbreak{}[\allowbreak{}5\allowbreak{}[\allowbreak{}1\allowbreak{}[\allowbreak{}2\allowbreak{}[\allowbreak{}7\allowbreak{}[\allowbreak{}4\allowbreak{}8\allowbreak{}]\allowbreak{}8\allowbreak{}]\allowbreak{}4\allowbreak{}]\allowbreak{}2\allowbreak{}[\allowbreak{}4\allowbreak{}]\allowbreak{}]\allowbreak{}]\allowbreak{}]\allowbreak{}]\allowbreak{}]\allowbreak{}5\allowbreak{}]\allowbreak{}]\allowbreak{}4\allowbreak{}7\allowbreak{}]\allowbreak{}7\allowbreak{}]\allowbreak{}]\allowbreak{}]\allowbreak{}]\allowbreak{}]\allowbreak{}5\allowbreak{}[\allowbreak{}6\allowbreak{}8\allowbreak{}[\allowbreak{}6\allowbreak{}[\allowbreak{}3\allowbreak{}[\allowbreak{}1\allowbreak{}]\allowbreak{}]\allowbreak{}]\allowbreak{}9\allowbreak{}[\allowbreak{}6\allowbreak{}]\allowbreak{}]\allowbreak{}]\allowbreak{}]\allowbreak{}5\allowbreak{}[\allowbreak{}6\allowbreak{}]\allowbreak{}}}\par}\smallskip

\paragraph{Unidirectional model.}
\noindent{$d=3$, $6$ edges:}\\[2pt]
{\small\raggedright{\texttt{213:\ 1\allowbreak{}3\allowbreak{}[\allowbreak{}1\allowbreak{}[\allowbreak{}2\allowbreak{}$^*$\allowbreak{}3\allowbreak{}[\allowbreak{}2\allowbreak{}]\allowbreak{}]\allowbreak{}]\allowbreak{}}}\par}\smallskip
\noindent{$d=4$, $12$ edges:}\\[2pt]
{\small\raggedright{\texttt{4321:\ 1\allowbreak{}[\allowbreak{}2\allowbreak{}[\allowbreak{}3\allowbreak{}[\allowbreak{}1\allowbreak{}[\allowbreak{}4\allowbreak{}]\allowbreak{}4\allowbreak{}$^*$\allowbreak{}]\allowbreak{}4\allowbreak{}[\allowbreak{}3\allowbreak{}]\allowbreak{}]\allowbreak{}3\allowbreak{}[\allowbreak{}2\allowbreak{}]\allowbreak{}]\allowbreak{}2\allowbreak{}[\allowbreak{}1\allowbreak{}]\allowbreak{}}}\par}\smallskip
\noindent{$d=5$, $30$ edges:}\\[2pt]
{\small\raggedright{\texttt{43215:\ 1\allowbreak{}[\allowbreak{}3\allowbreak{}5\allowbreak{}[\allowbreak{}3\allowbreak{}[\allowbreak{}1\allowbreak{}[\allowbreak{}2\allowbreak{}4\allowbreak{}[\allowbreak{}2\allowbreak{}[\allowbreak{}5\allowbreak{}[\allowbreak{}1\allowbreak{}3\allowbreak{}]\allowbreak{}]\allowbreak{}]\allowbreak{}5\allowbreak{}]\allowbreak{}2\allowbreak{}[\allowbreak{}1\allowbreak{}]\allowbreak{}]\allowbreak{}4\allowbreak{}]\allowbreak{}]\allowbreak{}2\allowbreak{}[\allowbreak{}1\allowbreak{}]\allowbreak{}5\allowbreak{}[\allowbreak{}1\allowbreak{}[\allowbreak{}2\allowbreak{}[\allowbreak{}3\allowbreak{}[\allowbreak{}4\allowbreak{}$^*$\allowbreak{}5\allowbreak{}[\allowbreak{}4\allowbreak{}]\allowbreak{}]\allowbreak{}4\allowbreak{}[\allowbreak{}3\allowbreak{}]\allowbreak{}5\allowbreak{}[\allowbreak{}4\allowbreak{}[\allowbreak{}3\allowbreak{}]\allowbreak{}]\allowbreak{}]\allowbreak{}]\allowbreak{}2\allowbreak{}]\allowbreak{}}}\par}\smallskip
\noindent{$d=6$, $60$ edges:}\\[2pt]
{\small\raggedright{\texttt{623145:\ 1\allowbreak{}[\allowbreak{}4\allowbreak{}[\allowbreak{}2\allowbreak{}[\allowbreak{}3\allowbreak{}5\allowbreak{}[\allowbreak{}3\allowbreak{}[\allowbreak{}6\allowbreak{}[\allowbreak{}1\allowbreak{}[\allowbreak{}2\allowbreak{}[\allowbreak{}3\allowbreak{}[\allowbreak{}4\allowbreak{}[\allowbreak{}5\allowbreak{}[\allowbreak{}6\allowbreak{}[\allowbreak{}2\allowbreak{}[\allowbreak{}3\allowbreak{}[\allowbreak{}5\allowbreak{}[\allowbreak{}1\allowbreak{}[\allowbreak{}6\allowbreak{}[\allowbreak{}3\allowbreak{}[\allowbreak{}4\allowbreak{}]\allowbreak{}]\allowbreak{}]\allowbreak{}4\allowbreak{}[\allowbreak{}1\allowbreak{}[\allowbreak{}6\allowbreak{}]\allowbreak{}]\allowbreak{}]\allowbreak{}]\allowbreak{}]\allowbreak{}]\allowbreak{}]\allowbreak{}6\allowbreak{}[\allowbreak{}2\allowbreak{}[\allowbreak{}5\allowbreak{}[\allowbreak{}1\allowbreak{}[\allowbreak{}3\allowbreak{}[\allowbreak{}4\allowbreak{}]\allowbreak{}]\allowbreak{}4\allowbreak{}]\allowbreak{}]\allowbreak{}5\allowbreak{}[\allowbreak{}1\allowbreak{}[\allowbreak{}2\allowbreak{}[\allowbreak{}4\allowbreak{}[\allowbreak{}6\allowbreak{}]\allowbreak{}6\allowbreak{}]\allowbreak{}]\allowbreak{}3\allowbreak{}]\allowbreak{}]\allowbreak{}]\allowbreak{}]\allowbreak{}]\allowbreak{}]\allowbreak{}2\allowbreak{}]\allowbreak{}]\allowbreak{}6\allowbreak{}]\allowbreak{}]\allowbreak{}]\allowbreak{}5\allowbreak{}[\allowbreak{}4\allowbreak{}[\allowbreak{}2\allowbreak{}[\allowbreak{}1\allowbreak{}]\allowbreak{}3\allowbreak{}[\allowbreak{}2\allowbreak{}[\allowbreak{}1\allowbreak{}[\allowbreak{}5\allowbreak{}]\allowbreak{}]\allowbreak{}]\allowbreak{}]\allowbreak{}]\allowbreak{}]\allowbreak{}4\allowbreak{}[\allowbreak{}1\allowbreak{}[\allowbreak{}3\allowbreak{}[\allowbreak{}2\allowbreak{}[\allowbreak{}6\allowbreak{}$^*$\allowbreak{}]\allowbreak{}5\allowbreak{}[\allowbreak{}2\allowbreak{}]\allowbreak{}6\allowbreak{}[\allowbreak{}5\allowbreak{}[\allowbreak{}4\allowbreak{}]\allowbreak{}]\allowbreak{}]\allowbreak{}5\allowbreak{}]\allowbreak{}3\allowbreak{}[\allowbreak{}1\allowbreak{}]\allowbreak{}]\allowbreak{}}}\par}\smallskip
\noindent{$d=7$, $140$ edges:}\\[2pt]
{\small\raggedright{\texttt{2345716:\ 1\allowbreak{}6\allowbreak{}[\allowbreak{}5\allowbreak{}[\allowbreak{}4\allowbreak{}[\allowbreak{}1\allowbreak{}[\allowbreak{}3\allowbreak{}[\allowbreak{}7\allowbreak{}[\allowbreak{}2\allowbreak{}[\allowbreak{}5\allowbreak{}[\allowbreak{}6\allowbreak{}[\allowbreak{}1\allowbreak{}[\allowbreak{}7\allowbreak{}]\allowbreak{}3\allowbreak{}]\allowbreak{}]\allowbreak{}]\allowbreak{}]\allowbreak{}]\allowbreak{}7\allowbreak{}[\allowbreak{}5\allowbreak{}[\allowbreak{}4\allowbreak{}[\allowbreak{}2\allowbreak{}[\allowbreak{}1\allowbreak{}[\allowbreak{}6\allowbreak{}[\allowbreak{}3\allowbreak{}[\allowbreak{}4\allowbreak{}[\allowbreak{}7\allowbreak{}[\allowbreak{}1\allowbreak{}]\allowbreak{}]\allowbreak{}7\allowbreak{}]\allowbreak{}]\allowbreak{}]\allowbreak{}6\allowbreak{}[\allowbreak{}7\allowbreak{}[\allowbreak{}3\allowbreak{}]\allowbreak{}]\allowbreak{}]\allowbreak{}3\allowbreak{}[\allowbreak{}2\allowbreak{}$^*$\allowbreak{}]\allowbreak{}]\allowbreak{}6\allowbreak{}[\allowbreak{}1\allowbreak{}[\allowbreak{}2\allowbreak{}[\allowbreak{}5\allowbreak{}[\allowbreak{}3\allowbreak{}[\allowbreak{}4\allowbreak{}6\allowbreak{}[\allowbreak{}1\allowbreak{}7\allowbreak{}[\allowbreak{}1\allowbreak{}[\allowbreak{}2\allowbreak{}[\allowbreak{}3\allowbreak{}[\allowbreak{}4\allowbreak{}[\allowbreak{}5\allowbreak{}[\allowbreak{}6\allowbreak{}[\allowbreak{}7\allowbreak{}]\allowbreak{}7\allowbreak{}[\allowbreak{}1\allowbreak{}]\allowbreak{}]\allowbreak{}]\allowbreak{}7\allowbreak{}[\allowbreak{}4\allowbreak{}[\allowbreak{}2\allowbreak{}[\allowbreak{}5\allowbreak{}]\allowbreak{}]\allowbreak{}6\allowbreak{}[\allowbreak{}2\allowbreak{}[\allowbreak{}4\allowbreak{}[\allowbreak{}5\allowbreak{}]\allowbreak{}]\allowbreak{}5\allowbreak{}[\allowbreak{}2\allowbreak{}]\allowbreak{}]\allowbreak{}]\allowbreak{}]\allowbreak{}]\allowbreak{}3\allowbreak{}[\allowbreak{}4\allowbreak{}[\allowbreak{}2\allowbreak{}[\allowbreak{}7\allowbreak{}[\allowbreak{}1\allowbreak{}[\allowbreak{}5\allowbreak{}[\allowbreak{}2\allowbreak{}3\allowbreak{}]\allowbreak{}]\allowbreak{}6\allowbreak{}]\allowbreak{}]\allowbreak{}]\allowbreak{}5\allowbreak{}6\allowbreak{}]\allowbreak{}]\allowbreak{}4\allowbreak{}[\allowbreak{}5\allowbreak{}]\allowbreak{}]\allowbreak{}]\allowbreak{}]\allowbreak{}4\allowbreak{}[\allowbreak{}7\allowbreak{}]\allowbreak{}6\allowbreak{}]\allowbreak{}7\allowbreak{}]\allowbreak{}4\allowbreak{}]\allowbreak{}]\allowbreak{}]\allowbreak{}]\allowbreak{}]\allowbreak{}3\allowbreak{}[\allowbreak{}1\allowbreak{}[\allowbreak{}5\allowbreak{}[\allowbreak{}4\allowbreak{}[\allowbreak{}2\allowbreak{}[\allowbreak{}3\allowbreak{}]\allowbreak{}]\allowbreak{}6\allowbreak{}]\allowbreak{}]\allowbreak{}]\allowbreak{}]\allowbreak{}7\allowbreak{}[\allowbreak{}3\allowbreak{}]\allowbreak{}]\allowbreak{}7\allowbreak{}[\allowbreak{}1\allowbreak{}5\allowbreak{}[\allowbreak{}4\allowbreak{}]\allowbreak{}]\allowbreak{}]\allowbreak{}}}\par}\smallskip
{\small\raggedright{\texttt{6475312:\ 1\allowbreak{}[\allowbreak{}3\allowbreak{}[\allowbreak{}2\allowbreak{}[\allowbreak{}5\allowbreak{}[\allowbreak{}1\allowbreak{}[\allowbreak{}7\allowbreak{}]\allowbreak{}]\allowbreak{}6\allowbreak{}[\allowbreak{}4\allowbreak{}]\allowbreak{}]\allowbreak{}5\allowbreak{}[\allowbreak{}2\allowbreak{}[\allowbreak{}7\allowbreak{}[\allowbreak{}1\allowbreak{}[\allowbreak{}4\allowbreak{}[\allowbreak{}2\allowbreak{}6\allowbreak{}[\allowbreak{}2\allowbreak{}[\allowbreak{}3\allowbreak{}]\allowbreak{}7\allowbreak{}]\allowbreak{}]\allowbreak{}6\allowbreak{}[\allowbreak{}4\allowbreak{}[\allowbreak{}5\allowbreak{}[\allowbreak{}2\allowbreak{}]\allowbreak{}]\allowbreak{}5\allowbreak{}[\allowbreak{}3\allowbreak{}[\allowbreak{}4\allowbreak{}]\allowbreak{}]\allowbreak{}]\allowbreak{}]\allowbreak{}3\allowbreak{}[\allowbreak{}1\allowbreak{}]\allowbreak{}4\allowbreak{}[\allowbreak{}3\allowbreak{}]\allowbreak{}]\allowbreak{}]\allowbreak{}7\allowbreak{}[\allowbreak{}2\allowbreak{}4\allowbreak{}[\allowbreak{}6\allowbreak{}$^*$\allowbreak{}]\allowbreak{}6\allowbreak{}]\allowbreak{}]\allowbreak{}]\allowbreak{}]\allowbreak{}2\allowbreak{}[\allowbreak{}1\allowbreak{}[\allowbreak{}4\allowbreak{}[\allowbreak{}3\allowbreak{}[\allowbreak{}6\allowbreak{}[\allowbreak{}1\allowbreak{}5\allowbreak{}[\allowbreak{}2\allowbreak{}]\allowbreak{}]\allowbreak{}]\allowbreak{}5\allowbreak{}[\allowbreak{}6\allowbreak{}]\allowbreak{}7\allowbreak{}[\allowbreak{}3\allowbreak{}[\allowbreak{}5\allowbreak{}[\allowbreak{}1\allowbreak{}]\allowbreak{}6\allowbreak{}]\allowbreak{}]\allowbreak{}]\allowbreak{}]\allowbreak{}]\allowbreak{}3\allowbreak{}[\allowbreak{}2\allowbreak{}[\allowbreak{}1\allowbreak{}]\allowbreak{}]\allowbreak{}}}\par}\smallskip
\noindent{$d=8$, $297$ edges:}\\[2pt]
{\small\raggedright{\texttt{12378456:\ 5\allowbreak{}[\allowbreak{}4\allowbreak{}[\allowbreak{}8\allowbreak{}[\allowbreak{}6\allowbreak{}]\allowbreak{}]\allowbreak{}6\allowbreak{}[\allowbreak{}7\allowbreak{}]\allowbreak{}8\allowbreak{}]\allowbreak{}6\allowbreak{}[\allowbreak{}4\allowbreak{}[\allowbreak{}5\allowbreak{}[\allowbreak{}3\allowbreak{}[\allowbreak{}8\allowbreak{}[\allowbreak{}2\allowbreak{}[\allowbreak{}6\allowbreak{}7\allowbreak{}]\allowbreak{}]\allowbreak{}]\allowbreak{}]\allowbreak{}]\allowbreak{}]\allowbreak{}8\allowbreak{}[\allowbreak{}6\allowbreak{}[\allowbreak{}5\allowbreak{}[\allowbreak{}4\allowbreak{}[\allowbreak{}1\allowbreak{}[\allowbreak{}2\allowbreak{}]\allowbreak{}2\allowbreak{}[\allowbreak{}7\allowbreak{}[\allowbreak{}1\allowbreak{}[\allowbreak{}5\allowbreak{}]\allowbreak{}6\allowbreak{}[\allowbreak{}4\allowbreak{}[\allowbreak{}1\allowbreak{}]\allowbreak{}]\allowbreak{}]\allowbreak{}]\allowbreak{}7\allowbreak{}[\allowbreak{}3\allowbreak{}[\allowbreak{}2\allowbreak{}[\allowbreak{}5\allowbreak{}]\allowbreak{}]\allowbreak{}]\allowbreak{}]\allowbreak{}7\allowbreak{}[\allowbreak{}4\allowbreak{}[\allowbreak{}2\allowbreak{}8\allowbreak{}[\allowbreak{}1\allowbreak{}[\allowbreak{}5\allowbreak{}[\allowbreak{}2\allowbreak{}[\allowbreak{}7\allowbreak{}]\allowbreak{}]\allowbreak{}]\allowbreak{}]\allowbreak{}]\allowbreak{}]\allowbreak{}]\allowbreak{}]\allowbreak{}7\allowbreak{}[\allowbreak{}3\allowbreak{}[\allowbreak{}2\allowbreak{}[\allowbreak{}1\allowbreak{}$^*$\allowbreak{}]\allowbreak{}]\allowbreak{}4\allowbreak{}[\allowbreak{}3\allowbreak{}6\allowbreak{}[\allowbreak{}3\allowbreak{}]\allowbreak{}]\allowbreak{}6\allowbreak{}[\allowbreak{}2\allowbreak{}3\allowbreak{}]\allowbreak{}]\allowbreak{}]\allowbreak{}}}\par}\smallskip
{\small\raggedright{\texttt{82435167:\ 1\allowbreak{}[\allowbreak{}5\allowbreak{}[\allowbreak{}3\allowbreak{}[\allowbreak{}2\allowbreak{}[\allowbreak{}7\allowbreak{}]\allowbreak{}4\allowbreak{}[\allowbreak{}2\allowbreak{}[\allowbreak{}8\allowbreak{}$^*$\allowbreak{}]\allowbreak{}]\allowbreak{}6\allowbreak{}]\allowbreak{}]\allowbreak{}]\allowbreak{}5\allowbreak{}[\allowbreak{}1\allowbreak{}[\allowbreak{}4\allowbreak{}6\allowbreak{}[\allowbreak{}4\allowbreak{}[\allowbreak{}8\allowbreak{}[\allowbreak{}7\allowbreak{}[\allowbreak{}2\allowbreak{}[\allowbreak{}3\allowbreak{}[\allowbreak{}4\allowbreak{}[\allowbreak{}1\allowbreak{}[\allowbreak{}7\allowbreak{}[\allowbreak{}5\allowbreak{}[\allowbreak{}8\allowbreak{}[\allowbreak{}3\allowbreak{}[\allowbreak{}6\allowbreak{}]\allowbreak{}6\allowbreak{}[\allowbreak{}1\allowbreak{}2\allowbreak{}[\allowbreak{}4\allowbreak{}[\allowbreak{}1\allowbreak{}[\allowbreak{}3\allowbreak{}[\allowbreak{}5\allowbreak{}[\allowbreak{}6\allowbreak{}[\allowbreak{}2\allowbreak{}[\allowbreak{}7\allowbreak{}]\allowbreak{}7\allowbreak{}[\allowbreak{}2\allowbreak{}[\allowbreak{}1\allowbreak{}[\allowbreak{}4\allowbreak{}]\allowbreak{}3\allowbreak{}[\allowbreak{}5\allowbreak{}[\allowbreak{}6\allowbreak{}[\allowbreak{}1\allowbreak{}[\allowbreak{}8\allowbreak{}[\allowbreak{}2\allowbreak{}[\allowbreak{}3\allowbreak{}[\allowbreak{}1\allowbreak{}[\allowbreak{}7\allowbreak{}]\allowbreak{}4\allowbreak{}[\allowbreak{}6\allowbreak{}7\allowbreak{}[\allowbreak{}6\allowbreak{}[\allowbreak{}2\allowbreak{}8\allowbreak{}[\allowbreak{}1\allowbreak{}[\allowbreak{}4\allowbreak{}[\allowbreak{}3\allowbreak{}5\allowbreak{}[\allowbreak{}3\allowbreak{}]\allowbreak{}]\allowbreak{}]\allowbreak{}]\allowbreak{}]\allowbreak{}8\allowbreak{}]\allowbreak{}]\allowbreak{}5\allowbreak{}[\allowbreak{}1\allowbreak{}]\allowbreak{}]\allowbreak{}]\allowbreak{}3\allowbreak{}[\allowbreak{}4\allowbreak{}]\allowbreak{}7\allowbreak{}[\allowbreak{}5\allowbreak{}[\allowbreak{}6\allowbreak{}[\allowbreak{}3\allowbreak{}[\allowbreak{}4\allowbreak{}8\allowbreak{}[\allowbreak{}5\allowbreak{}[\allowbreak{}4\allowbreak{}7\allowbreak{}[\allowbreak{}1\allowbreak{}[\allowbreak{}2\allowbreak{}[\allowbreak{}3\allowbreak{}[\allowbreak{}4\allowbreak{}[\allowbreak{}5\allowbreak{}[\allowbreak{}6\allowbreak{}[\allowbreak{}7\allowbreak{}[\allowbreak{}1\allowbreak{}[\allowbreak{}4\allowbreak{}8\allowbreak{}]\allowbreak{}8\allowbreak{}]\allowbreak{}]\allowbreak{}]\allowbreak{}]\allowbreak{}]\allowbreak{}6\allowbreak{}[\allowbreak{}5\allowbreak{}[\allowbreak{}3\allowbreak{}4\allowbreak{}[\allowbreak{}8\allowbreak{}[\allowbreak{}3\allowbreak{}[\allowbreak{}1\allowbreak{}[\allowbreak{}2\allowbreak{}5\allowbreak{}6\allowbreak{}[\allowbreak{}2\allowbreak{}[\allowbreak{}7\allowbreak{}]\allowbreak{}5\allowbreak{}7\allowbreak{}[\allowbreak{}5\allowbreak{}]\allowbreak{}]\allowbreak{}]\allowbreak{}7\allowbreak{}[\allowbreak{}1\allowbreak{}[\allowbreak{}5\allowbreak{}]\allowbreak{}]\allowbreak{}]\allowbreak{}]\allowbreak{}]\allowbreak{}8\allowbreak{}[\allowbreak{}3\allowbreak{}[\allowbreak{}1\allowbreak{}]\allowbreak{}]\allowbreak{}]\allowbreak{}8\allowbreak{}[\allowbreak{}7\allowbreak{}[\allowbreak{}4\allowbreak{}[\allowbreak{}5\allowbreak{}[\allowbreak{}6\allowbreak{}]\allowbreak{}]\allowbreak{}5\allowbreak{}]\allowbreak{}]\allowbreak{}]\allowbreak{}8\allowbreak{}]\allowbreak{}4\allowbreak{}[\allowbreak{}2\allowbreak{}[\allowbreak{}8\allowbreak{}[\allowbreak{}3\allowbreak{}]\allowbreak{}]\allowbreak{}]\allowbreak{}]\allowbreak{}]\allowbreak{}]\allowbreak{}]\allowbreak{}]\allowbreak{}]\allowbreak{}]\allowbreak{}]\allowbreak{}]\allowbreak{}]\allowbreak{}]\allowbreak{}]\allowbreak{}8\allowbreak{}]\allowbreak{}4\allowbreak{}[\allowbreak{}3\allowbreak{}8\allowbreak{}[\allowbreak{}6\allowbreak{}[\allowbreak{}3\allowbreak{}[\allowbreak{}5\allowbreak{}[\allowbreak{}2\allowbreak{}[\allowbreak{}8\allowbreak{}[\allowbreak{}4\allowbreak{}]\allowbreak{}]\allowbreak{}]\allowbreak{}7\allowbreak{}[\allowbreak{}5\allowbreak{}[\allowbreak{}2\allowbreak{}[\allowbreak{}8\allowbreak{}]\allowbreak{}]\allowbreak{}]\allowbreak{}]\allowbreak{}5\allowbreak{}]\allowbreak{}]\allowbreak{}]\allowbreak{}8\allowbreak{}[\allowbreak{}6\allowbreak{}[\allowbreak{}3\allowbreak{}]\allowbreak{}]\allowbreak{}]\allowbreak{}]\allowbreak{}]\allowbreak{}]\allowbreak{}]\allowbreak{}]\allowbreak{}]\allowbreak{}]\allowbreak{}]\allowbreak{}]\allowbreak{}]\allowbreak{}]\allowbreak{}8\allowbreak{}[\allowbreak{}7\allowbreak{}[\allowbreak{}2\allowbreak{}[\allowbreak{}5\allowbreak{}[\allowbreak{}1\allowbreak{}[\allowbreak{}4\allowbreak{}[\allowbreak{}6\allowbreak{}[\allowbreak{}7\allowbreak{}[\allowbreak{}5\allowbreak{}]\allowbreak{}]\allowbreak{}]\allowbreak{}]\allowbreak{}4\allowbreak{}]\allowbreak{}]\allowbreak{}]\allowbreak{}]\allowbreak{}]\allowbreak{}]\allowbreak{}5\allowbreak{}6\allowbreak{}[\allowbreak{}1\allowbreak{}[\allowbreak{}8\allowbreak{}]\allowbreak{}]\allowbreak{}]\allowbreak{}4\allowbreak{}[\allowbreak{}5\allowbreak{}[\allowbreak{}3\allowbreak{}[\allowbreak{}1\allowbreak{}[\allowbreak{}7\allowbreak{}8\allowbreak{}[\allowbreak{}7\allowbreak{}]\allowbreak{}]\allowbreak{}7\allowbreak{}[\allowbreak{}6\allowbreak{}[\allowbreak{}1\allowbreak{}]\allowbreak{}8\allowbreak{}[\allowbreak{}5\allowbreak{}[\allowbreak{}2\allowbreak{}[\allowbreak{}6\allowbreak{}[\allowbreak{}1\allowbreak{}]\allowbreak{}]\allowbreak{}4\allowbreak{}[\allowbreak{}2\allowbreak{}[\allowbreak{}6\allowbreak{}[\allowbreak{}3\allowbreak{}]\allowbreak{}]\allowbreak{}3\allowbreak{}[\allowbreak{}6\allowbreak{}[\allowbreak{}1\allowbreak{}2\allowbreak{}[\allowbreak{}7\allowbreak{}8\allowbreak{}[\allowbreak{}1\allowbreak{}[\allowbreak{}4\allowbreak{}[\allowbreak{}6\allowbreak{}]\allowbreak{}]\allowbreak{}7\allowbreak{}[\allowbreak{}1\allowbreak{}]\allowbreak{}]\allowbreak{}]\allowbreak{}]\allowbreak{}]\allowbreak{}]\allowbreak{}]\allowbreak{}]\allowbreak{}]\allowbreak{}]\allowbreak{}8\allowbreak{}[\allowbreak{}1\allowbreak{}[\allowbreak{}7\allowbreak{}]\allowbreak{}7\allowbreak{}]\allowbreak{}]\allowbreak{}]\allowbreak{}]\allowbreak{}]\allowbreak{}]\allowbreak{}]\allowbreak{}]\allowbreak{}]\allowbreak{}3\allowbreak{}]\allowbreak{}}}\par}\smallskip
{\small\raggedright{\texttt{63715428:\ 2\allowbreak{}[\allowbreak{}4\allowbreak{}[\allowbreak{}1\allowbreak{}5\allowbreak{}[\allowbreak{}1\allowbreak{}[\allowbreak{}7\allowbreak{}[\allowbreak{}3\allowbreak{}[\allowbreak{}2\allowbreak{}6\allowbreak{}$^*$\allowbreak{}8\allowbreak{}[\allowbreak{}4\allowbreak{}[\allowbreak{}2\allowbreak{}5\allowbreak{}6\allowbreak{}[\allowbreak{}2\allowbreak{}]\allowbreak{}]\allowbreak{}6\allowbreak{}[\allowbreak{}1\allowbreak{}]\allowbreak{}]\allowbreak{}]\allowbreak{}]\allowbreak{}]\allowbreak{}]\allowbreak{}]\allowbreak{}]\allowbreak{}4\allowbreak{}8\allowbreak{}[\allowbreak{}2\allowbreak{}[\allowbreak{}4\allowbreak{}[\allowbreak{}7\allowbreak{}[\allowbreak{}1\allowbreak{}[\allowbreak{}3\allowbreak{}6\allowbreak{}[\allowbreak{}2\allowbreak{}[\allowbreak{}3\allowbreak{}[\allowbreak{}5\allowbreak{}[\allowbreak{}4\allowbreak{}]\allowbreak{}]\allowbreak{}]\allowbreak{}3\allowbreak{}8\allowbreak{}]\allowbreak{}]\allowbreak{}]\allowbreak{}]\allowbreak{}5\allowbreak{}[\allowbreak{}7\allowbreak{}[\allowbreak{}3\allowbreak{}[\allowbreak{}1\allowbreak{}[\allowbreak{}4\allowbreak{}[\allowbreak{}2\allowbreak{}[\allowbreak{}6\allowbreak{}]\allowbreak{}6\allowbreak{}[\allowbreak{}2\allowbreak{}[\allowbreak{}7\allowbreak{}[\allowbreak{}5\allowbreak{}[\allowbreak{}6\allowbreak{}]\allowbreak{}]\allowbreak{}8\allowbreak{}]\allowbreak{}3\allowbreak{}[\allowbreak{}5\allowbreak{}]\allowbreak{}]\allowbreak{}]\allowbreak{}]\allowbreak{}4\allowbreak{}]\allowbreak{}]\allowbreak{}]\allowbreak{}]\allowbreak{}4\allowbreak{}[\allowbreak{}1\allowbreak{}[\allowbreak{}2\allowbreak{}[\allowbreak{}5\allowbreak{}[\allowbreak{}3\allowbreak{}]\allowbreak{}7\allowbreak{}]\allowbreak{}]\allowbreak{}]\allowbreak{}]\allowbreak{}}}\par}\smallskip
{\small\raggedright{\texttt{15743628:\ 6\allowbreak{}[\allowbreak{}2\allowbreak{}3\allowbreak{}[\allowbreak{}4\allowbreak{}[\allowbreak{}7\allowbreak{}[\allowbreak{}1\allowbreak{}[\allowbreak{}8\allowbreak{}]\allowbreak{}5\allowbreak{}[\allowbreak{}1\allowbreak{}$^*$\allowbreak{}]\allowbreak{}]\allowbreak{}8\allowbreak{}]\allowbreak{}]\allowbreak{}]\allowbreak{}}}\par}\smallskip
\noindent{$d=9$, $656$ edges:}\\[2pt]
{\small\raggedright{\texttt{987654321:\ 1\allowbreak{}[\allowbreak{}2\allowbreak{}[\allowbreak{}3\allowbreak{}[\allowbreak{}4\allowbreak{}[\allowbreak{}5\allowbreak{}[\allowbreak{}6\allowbreak{}[\allowbreak{}7\allowbreak{}[\allowbreak{}2\allowbreak{}8\allowbreak{}[\allowbreak{}1\allowbreak{}[\allowbreak{}2\allowbreak{}[\allowbreak{}5\allowbreak{}[\allowbreak{}9\allowbreak{}[\allowbreak{}3\allowbreak{}]\allowbreak{}]\allowbreak{}9\allowbreak{}[\allowbreak{}5\allowbreak{}[\allowbreak{}4\allowbreak{}[\allowbreak{}3\allowbreak{}]\allowbreak{}]\allowbreak{}7\allowbreak{}[\allowbreak{}8\allowbreak{}]\allowbreak{}]\allowbreak{}]\allowbreak{}3\allowbreak{}[\allowbreak{}2\allowbreak{}[\allowbreak{}6\allowbreak{}[\allowbreak{}9\allowbreak{}]\allowbreak{}9\allowbreak{}[\allowbreak{}6\allowbreak{}[\allowbreak{}8\allowbreak{}[\allowbreak{}4\allowbreak{}]\allowbreak{}]\allowbreak{}]\allowbreak{}]\allowbreak{}4\allowbreak{}[\allowbreak{}6\allowbreak{}]\allowbreak{}5\allowbreak{}[\allowbreak{}6\allowbreak{}[\allowbreak{}4\allowbreak{}]\allowbreak{}]\allowbreak{}9\allowbreak{}[\allowbreak{}5\allowbreak{}[\allowbreak{}2\allowbreak{}4\allowbreak{}[\allowbreak{}2\allowbreak{}[\allowbreak{}1\allowbreak{}[\allowbreak{}7\allowbreak{}[\allowbreak{}9\allowbreak{}[\allowbreak{}5\allowbreak{}]\allowbreak{}]\allowbreak{}]\allowbreak{}]\allowbreak{}6\allowbreak{}]\allowbreak{}6\allowbreak{}[\allowbreak{}2\allowbreak{}[\allowbreak{}7\allowbreak{}[\allowbreak{}3\allowbreak{}[\allowbreak{}1\allowbreak{}[\allowbreak{}5\allowbreak{}[\allowbreak{}4\allowbreak{}[\allowbreak{}9\allowbreak{}[\allowbreak{}7\allowbreak{}[\allowbreak{}1\allowbreak{}2\allowbreak{}[\allowbreak{}1\allowbreak{}5\allowbreak{}[\allowbreak{}3\allowbreak{}4\allowbreak{}6\allowbreak{}]\allowbreak{}]\allowbreak{}8\allowbreak{}]\allowbreak{}]\allowbreak{}]\allowbreak{}]\allowbreak{}]\allowbreak{}]\allowbreak{}4\allowbreak{}[\allowbreak{}1\allowbreak{}3\allowbreak{}[\allowbreak{}1\allowbreak{}[\allowbreak{}5\allowbreak{}]\allowbreak{}8\allowbreak{}]\allowbreak{}8\allowbreak{}[\allowbreak{}6\allowbreak{}[\allowbreak{}9\allowbreak{}]\allowbreak{}]\allowbreak{}]\allowbreak{}8\allowbreak{}]\allowbreak{}8\allowbreak{}[\allowbreak{}7\allowbreak{}[\allowbreak{}1\allowbreak{}3\allowbreak{}[\allowbreak{}4\allowbreak{}[\allowbreak{}5\allowbreak{}[\allowbreak{}8\allowbreak{}]\allowbreak{}]\allowbreak{}]\allowbreak{}]\allowbreak{}]\allowbreak{}]\allowbreak{}]\allowbreak{}]\allowbreak{}6\allowbreak{}[\allowbreak{}5\allowbreak{}]\allowbreak{}]\allowbreak{}]\allowbreak{}4\allowbreak{}[\allowbreak{}2\allowbreak{}3\allowbreak{}]\allowbreak{}9\allowbreak{}]\allowbreak{}9\allowbreak{}$^*$\allowbreak{}]\allowbreak{}]\allowbreak{}9\allowbreak{}[\allowbreak{}1\allowbreak{}[\allowbreak{}2\allowbreak{}]\allowbreak{}8\allowbreak{}[\allowbreak{}3\allowbreak{}7\allowbreak{}]\allowbreak{}]\allowbreak{}]\allowbreak{}7\allowbreak{}[\allowbreak{}1\allowbreak{}6\allowbreak{}[\allowbreak{}3\allowbreak{}9\allowbreak{}[\allowbreak{}2\allowbreak{}8\allowbreak{}]\allowbreak{}]\allowbreak{}9\allowbreak{}[\allowbreak{}1\allowbreak{}3\allowbreak{}6\allowbreak{}[\allowbreak{}3\allowbreak{}[\allowbreak{}2\allowbreak{}[\allowbreak{}4\allowbreak{}]\allowbreak{}]\allowbreak{}]\allowbreak{}]\allowbreak{}]\allowbreak{}9\allowbreak{}]\allowbreak{}]\allowbreak{}7\allowbreak{}[\allowbreak{}6\allowbreak{}[\allowbreak{}4\allowbreak{}[\allowbreak{}9\allowbreak{}]\allowbreak{}]\allowbreak{}]\allowbreak{}]\allowbreak{}]\allowbreak{}3\allowbreak{}[\allowbreak{}2\allowbreak{}[\allowbreak{}5\allowbreak{}[\allowbreak{}4\allowbreak{}[\allowbreak{}8\allowbreak{}]\allowbreak{}6\allowbreak{}[\allowbreak{}4\allowbreak{}[\allowbreak{}8\allowbreak{}]\allowbreak{}7\allowbreak{}[\allowbreak{}4\allowbreak{}]\allowbreak{}]\allowbreak{}7\allowbreak{}[\allowbreak{}4\allowbreak{}[\allowbreak{}8\allowbreak{}]\allowbreak{}8\allowbreak{}]\allowbreak{}]\allowbreak{}]\allowbreak{}4\allowbreak{}[\allowbreak{}2\allowbreak{}[\allowbreak{}9\allowbreak{}]\allowbreak{}6\allowbreak{}]\allowbreak{}]\allowbreak{}4\allowbreak{}[\allowbreak{}2\allowbreak{}3\allowbreak{}[\allowbreak{}7\allowbreak{}[\allowbreak{}2\allowbreak{}[\allowbreak{}5\allowbreak{}]\allowbreak{}]\allowbreak{}]\allowbreak{}]\allowbreak{}5\allowbreak{}[\allowbreak{}2\allowbreak{}3\allowbreak{}4\allowbreak{}]\allowbreak{}]\allowbreak{}2\allowbreak{}[\allowbreak{}1\allowbreak{}[\allowbreak{}4\allowbreak{}[\allowbreak{}3\allowbreak{}[\allowbreak{}6\allowbreak{}[\allowbreak{}5\allowbreak{}[\allowbreak{}8\allowbreak{}[\allowbreak{}2\allowbreak{}[\allowbreak{}9\allowbreak{}[\allowbreak{}3\allowbreak{}[\allowbreak{}4\allowbreak{}[\allowbreak{}1\allowbreak{}]\allowbreak{}7\allowbreak{}[\allowbreak{}5\allowbreak{}]\allowbreak{}]\allowbreak{}4\allowbreak{}[\allowbreak{}1\allowbreak{}[\allowbreak{}5\allowbreak{}[\allowbreak{}3\allowbreak{}[\allowbreak{}6\allowbreak{}]\allowbreak{}6\allowbreak{}[\allowbreak{}7\allowbreak{}[\allowbreak{}3\allowbreak{}[\allowbreak{}1\allowbreak{}]\allowbreak{}8\allowbreak{}[\allowbreak{}3\allowbreak{}]\allowbreak{}]\allowbreak{}]\allowbreak{}]\allowbreak{}6\allowbreak{}[\allowbreak{}3\allowbreak{}[\allowbreak{}2\allowbreak{}]\allowbreak{}7\allowbreak{}[\allowbreak{}3\allowbreak{}[\allowbreak{}2\allowbreak{}[\allowbreak{}5\allowbreak{}8\allowbreak{}[\allowbreak{}1\allowbreak{}[\allowbreak{}4\allowbreak{}]\allowbreak{}5\allowbreak{}[\allowbreak{}1\allowbreak{}[\allowbreak{}6\allowbreak{}]\allowbreak{}]\allowbreak{}]\allowbreak{}]\allowbreak{}5\allowbreak{}[\allowbreak{}4\allowbreak{}]\allowbreak{}8\allowbreak{}[\allowbreak{}1\allowbreak{}5\allowbreak{}]\allowbreak{}9\allowbreak{}[\allowbreak{}5\allowbreak{}8\allowbreak{}]\allowbreak{}]\allowbreak{}]\allowbreak{}]\allowbreak{}7\allowbreak{}[\allowbreak{}3\allowbreak{}[\allowbreak{}6\allowbreak{}]\allowbreak{}5\allowbreak{}[\allowbreak{}3\allowbreak{}[\allowbreak{}2\allowbreak{}[\allowbreak{}1\allowbreak{}6\allowbreak{}9\allowbreak{}[\allowbreak{}1\allowbreak{}[\allowbreak{}8\allowbreak{}]\allowbreak{}8\allowbreak{}[\allowbreak{}6\allowbreak{}[\allowbreak{}1\allowbreak{}[\allowbreak{}4\allowbreak{}[\allowbreak{}7\allowbreak{}]\allowbreak{}5\allowbreak{}[\allowbreak{}4\allowbreak{}]\allowbreak{}7\allowbreak{}]\allowbreak{}7\allowbreak{}[\allowbreak{}3\allowbreak{}[\allowbreak{}4\allowbreak{}]\allowbreak{}4\allowbreak{}[\allowbreak{}1\allowbreak{}]\allowbreak{}]\allowbreak{}]\allowbreak{}]\allowbreak{}]\allowbreak{}]\allowbreak{}6\allowbreak{}[\allowbreak{}2\allowbreak{}[\allowbreak{}8\allowbreak{}[\allowbreak{}1\allowbreak{}[\allowbreak{}4\allowbreak{}7\allowbreak{}[\allowbreak{}4\allowbreak{}[\allowbreak{}3\allowbreak{}9\allowbreak{}[\allowbreak{}5\allowbreak{}]\allowbreak{}]\allowbreak{}]\allowbreak{}]\allowbreak{}9\allowbreak{}[\allowbreak{}5\allowbreak{}[\allowbreak{}1\allowbreak{}]\allowbreak{}]\allowbreak{}]\allowbreak{}9\allowbreak{}[\allowbreak{}1\allowbreak{}[\allowbreak{}7\allowbreak{}[\allowbreak{}6\allowbreak{}[\allowbreak{}5\allowbreak{}]\allowbreak{}]\allowbreak{}]\allowbreak{}]\allowbreak{}]\allowbreak{}8\allowbreak{}[\allowbreak{}9\allowbreak{}[\allowbreak{}2\allowbreak{}]\allowbreak{}]\allowbreak{}9\allowbreak{}[\allowbreak{}1\allowbreak{}8\allowbreak{}[\allowbreak{}1\allowbreak{}5\allowbreak{}[\allowbreak{}1\allowbreak{}]\allowbreak{}7\allowbreak{}]\allowbreak{}]\allowbreak{}]\allowbreak{}8\allowbreak{}[\allowbreak{}2\allowbreak{}[\allowbreak{}4\allowbreak{}[\allowbreak{}1\allowbreak{}[\allowbreak{}5\allowbreak{}[\allowbreak{}7\allowbreak{}8\allowbreak{}9\allowbreak{}[\allowbreak{}2\allowbreak{}[\allowbreak{}6\allowbreak{}[\allowbreak{}4\allowbreak{}[\allowbreak{}8\allowbreak{}]\allowbreak{}]\allowbreak{}]\allowbreak{}8\allowbreak{}]\allowbreak{}]\allowbreak{}6\allowbreak{}[\allowbreak{}3\allowbreak{}[\allowbreak{}7\allowbreak{}]\allowbreak{}]\allowbreak{}7\allowbreak{}9\allowbreak{}]\allowbreak{}6\allowbreak{}[\allowbreak{}9\allowbreak{}[\allowbreak{}3\allowbreak{}[\allowbreak{}1\allowbreak{}[\allowbreak{}7\allowbreak{}[\allowbreak{}2\allowbreak{}[\allowbreak{}8\allowbreak{}]\allowbreak{}]\allowbreak{}8\allowbreak{}[\allowbreak{}7\allowbreak{}]\allowbreak{}]\allowbreak{}7\allowbreak{}8\allowbreak{}]\allowbreak{}]\allowbreak{}]\allowbreak{}9\allowbreak{}[\allowbreak{}6\allowbreak{}[\allowbreak{}1\allowbreak{}[\allowbreak{}5\allowbreak{}]\allowbreak{}5\allowbreak{}[\allowbreak{}3\allowbreak{}[\allowbreak{}4\allowbreak{}[\allowbreak{}2\allowbreak{}]\allowbreak{}7\allowbreak{}]\allowbreak{}7\allowbreak{}[\allowbreak{}1\allowbreak{}[\allowbreak{}2\allowbreak{}[\allowbreak{}8\allowbreak{}[\allowbreak{}6\allowbreak{}[\allowbreak{}3\allowbreak{}5\allowbreak{}[\allowbreak{}9\allowbreak{}[\allowbreak{}2\allowbreak{}]\allowbreak{}]\allowbreak{}]\allowbreak{}]\allowbreak{}]\allowbreak{}4\allowbreak{}]\allowbreak{}]\allowbreak{}]\allowbreak{}7\allowbreak{}[\allowbreak{}5\allowbreak{}[\allowbreak{}4\allowbreak{}]\allowbreak{}]\allowbreak{}]\allowbreak{}]\allowbreak{}]\allowbreak{}6\allowbreak{}[\allowbreak{}4\allowbreak{}[\allowbreak{}7\allowbreak{}[\allowbreak{}9\allowbreak{}]\allowbreak{}]\allowbreak{}]\allowbreak{}]\allowbreak{}6\allowbreak{}[\allowbreak{}1\allowbreak{}[\allowbreak{}2\allowbreak{}[\allowbreak{}9\allowbreak{}[\allowbreak{}5\allowbreak{}]\allowbreak{}]\allowbreak{}]\allowbreak{}4\allowbreak{}[\allowbreak{}7\allowbreak{}[\allowbreak{}5\allowbreak{}[\allowbreak{}1\allowbreak{}]\allowbreak{}]\allowbreak{}9\allowbreak{}[\allowbreak{}2\allowbreak{}]\allowbreak{}]\allowbreak{}]\allowbreak{}]\allowbreak{}9\allowbreak{}]\allowbreak{}6\allowbreak{}8\allowbreak{}[\allowbreak{}3\allowbreak{}[\allowbreak{}4\allowbreak{}[\allowbreak{}1\allowbreak{}2\allowbreak{}[\allowbreak{}5\allowbreak{}7\allowbreak{}[\allowbreak{}6\allowbreak{}[\allowbreak{}5\allowbreak{}[\allowbreak{}1\allowbreak{}]\allowbreak{}]\allowbreak{}]\allowbreak{}]\allowbreak{}6\allowbreak{}]\allowbreak{}9\allowbreak{}[\allowbreak{}2\allowbreak{}[\allowbreak{}5\allowbreak{}]\allowbreak{}]\allowbreak{}]\allowbreak{}6\allowbreak{}]\allowbreak{}]\allowbreak{}6\allowbreak{}[\allowbreak{}2\allowbreak{}[\allowbreak{}9\allowbreak{}[\allowbreak{}3\allowbreak{}[\allowbreak{}5\allowbreak{}[\allowbreak{}4\allowbreak{}[\allowbreak{}1\allowbreak{}[\allowbreak{}2\allowbreak{}7\allowbreak{}[\allowbreak{}9\allowbreak{}[\allowbreak{}4\allowbreak{}[\allowbreak{}2\allowbreak{}[\allowbreak{}3\allowbreak{}8\allowbreak{}]\allowbreak{}]\allowbreak{}]\allowbreak{}]\allowbreak{}8\allowbreak{}]\allowbreak{}7\allowbreak{}[\allowbreak{}2\allowbreak{}[\allowbreak{}9\allowbreak{}]\allowbreak{}]\allowbreak{}]\allowbreak{}6\allowbreak{}7\allowbreak{}[\allowbreak{}4\allowbreak{}]\allowbreak{}]\allowbreak{}]\allowbreak{}5\allowbreak{}]\allowbreak{}]\allowbreak{}8\allowbreak{}[\allowbreak{}5\allowbreak{}[\allowbreak{}4\allowbreak{}[\allowbreak{}9\allowbreak{}]\allowbreak{}]\allowbreak{}]\allowbreak{}]\allowbreak{}8\allowbreak{}[\allowbreak{}5\allowbreak{}[\allowbreak{}4\allowbreak{}[\allowbreak{}1\allowbreak{}2\allowbreak{}6\allowbreak{}]\allowbreak{}9\allowbreak{}[\allowbreak{}4\allowbreak{}[\allowbreak{}1\allowbreak{}6\allowbreak{}]\allowbreak{}]\allowbreak{}]\allowbreak{}]\allowbreak{}]\allowbreak{}]\allowbreak{}3\allowbreak{}[\allowbreak{}7\allowbreak{}]\allowbreak{}]\allowbreak{}7\allowbreak{}[\allowbreak{}3\allowbreak{}[\allowbreak{}1\allowbreak{}[\allowbreak{}4\allowbreak{}6\allowbreak{}[\allowbreak{}4\allowbreak{}[\allowbreak{}9\allowbreak{}[\allowbreak{}2\allowbreak{}]\allowbreak{}]\allowbreak{}9\allowbreak{}[\allowbreak{}2\allowbreak{}[\allowbreak{}4\allowbreak{}[\allowbreak{}7\allowbreak{}]\allowbreak{}]\allowbreak{}]\allowbreak{}]\allowbreak{}8\allowbreak{}[\allowbreak{}6\allowbreak{}[\allowbreak{}9\allowbreak{}]\allowbreak{}]\allowbreak{}]\allowbreak{}4\allowbreak{}]\allowbreak{}4\allowbreak{}[\allowbreak{}6\allowbreak{}[\allowbreak{}1\allowbreak{}[\allowbreak{}5\allowbreak{}[\allowbreak{}2\allowbreak{}[\allowbreak{}3\allowbreak{}4\allowbreak{}[\allowbreak{}3\allowbreak{}[\allowbreak{}1\allowbreak{}]\allowbreak{}]\allowbreak{}8\allowbreak{}[\allowbreak{}3\allowbreak{}[\allowbreak{}7\allowbreak{}[\allowbreak{}9\allowbreak{}]\allowbreak{}9\allowbreak{}[\allowbreak{}1\allowbreak{}]\allowbreak{}]\allowbreak{}4\allowbreak{}7\allowbreak{}[\allowbreak{}3\allowbreak{}4\allowbreak{}[\allowbreak{}3\allowbreak{}[\allowbreak{}1\allowbreak{}[\allowbreak{}9\allowbreak{}[\allowbreak{}5\allowbreak{}]\allowbreak{}]\allowbreak{}9\allowbreak{}[\allowbreak{}2\allowbreak{}[\allowbreak{}6\allowbreak{}[\allowbreak{}7\allowbreak{}[\allowbreak{}4\allowbreak{}]\allowbreak{}]\allowbreak{}]\allowbreak{}]\allowbreak{}]\allowbreak{}9\allowbreak{}[\allowbreak{}3\allowbreak{}[\allowbreak{}6\allowbreak{}]\allowbreak{}]\allowbreak{}]\allowbreak{}]\allowbreak{}]\allowbreak{}9\allowbreak{}[\allowbreak{}3\allowbreak{}]\allowbreak{}]\allowbreak{}3\allowbreak{}[\allowbreak{}2\allowbreak{}[\allowbreak{}9\allowbreak{}]\allowbreak{}7\allowbreak{}8\allowbreak{}9\allowbreak{}[\allowbreak{}7\allowbreak{}[\allowbreak{}2\allowbreak{}[\allowbreak{}6\allowbreak{}[\allowbreak{}1\allowbreak{}[\allowbreak{}4\allowbreak{}[\allowbreak{}5\allowbreak{}8\allowbreak{}[\allowbreak{}3\allowbreak{}[\allowbreak{}2\allowbreak{}5\allowbreak{}7\allowbreak{}9\allowbreak{}[\allowbreak{}5\allowbreak{}]\allowbreak{}]\allowbreak{}5\allowbreak{}[\allowbreak{}3\allowbreak{}[\allowbreak{}7\allowbreak{}[\allowbreak{}1\allowbreak{}]\allowbreak{}9\allowbreak{}[\allowbreak{}1\allowbreak{}]\allowbreak{}]\allowbreak{}7\allowbreak{}[\allowbreak{}9\allowbreak{}[\allowbreak{}6\allowbreak{}]\allowbreak{}]\allowbreak{}9\allowbreak{}]\allowbreak{}9\allowbreak{}[\allowbreak{}3\allowbreak{}5\allowbreak{}[\allowbreak{}2\allowbreak{}3\allowbreak{}7\allowbreak{}[\allowbreak{}2\allowbreak{}3\allowbreak{}[\allowbreak{}1\allowbreak{}]\allowbreak{}]\allowbreak{}]\allowbreak{}]\allowbreak{}]\allowbreak{}]\allowbreak{}5\allowbreak{}8\allowbreak{}]\allowbreak{}8\allowbreak{}[\allowbreak{}5\allowbreak{}[\allowbreak{}3\allowbreak{}4\allowbreak{}[\allowbreak{}7\allowbreak{}[\allowbreak{}3\allowbreak{}]\allowbreak{}]\allowbreak{}]\allowbreak{}]\allowbreak{}]\allowbreak{}8\allowbreak{}[\allowbreak{}5\allowbreak{}[\allowbreak{}6\allowbreak{}[\allowbreak{}1\allowbreak{}[\allowbreak{}9\allowbreak{}[\allowbreak{}7\allowbreak{}[\allowbreak{}2\allowbreak{}[\allowbreak{}3\allowbreak{}]\allowbreak{}]\allowbreak{}]\allowbreak{}]\allowbreak{}]\allowbreak{}]\allowbreak{}6\allowbreak{}[\allowbreak{}4\allowbreak{}]\allowbreak{}]\allowbreak{}]\allowbreak{}6\allowbreak{}[\allowbreak{}1\allowbreak{}[\allowbreak{}4\allowbreak{}]\allowbreak{}4\allowbreak{}[\allowbreak{}2\allowbreak{}]\allowbreak{}]\allowbreak{}8\allowbreak{}[\allowbreak{}4\allowbreak{}[\allowbreak{}6\allowbreak{}[\allowbreak{}2\allowbreak{}[\allowbreak{}1\allowbreak{}3\allowbreak{}[\allowbreak{}1\allowbreak{}[\allowbreak{}7\allowbreak{}[\allowbreak{}9\allowbreak{}[\allowbreak{}5\allowbreak{}[\allowbreak{}4\allowbreak{}]\allowbreak{}]\allowbreak{}]\allowbreak{}8\allowbreak{}]\allowbreak{}5\allowbreak{}[\allowbreak{}1\allowbreak{}[\allowbreak{}8\allowbreak{}[\allowbreak{}4\allowbreak{}[\allowbreak{}2\allowbreak{}[\allowbreak{}3\allowbreak{}]\allowbreak{}6\allowbreak{}9\allowbreak{}[\allowbreak{}6\allowbreak{}]\allowbreak{}]\allowbreak{}]\allowbreak{}]\allowbreak{}]\allowbreak{}]\allowbreak{}5\allowbreak{}[\allowbreak{}1\allowbreak{}[\allowbreak{}7\allowbreak{}[\allowbreak{}3\allowbreak{}4\allowbreak{}[\allowbreak{}2\allowbreak{}[\allowbreak{}3\allowbreak{}8\allowbreak{}]\allowbreak{}]\allowbreak{}]\allowbreak{}]\allowbreak{}3\allowbreak{}7\allowbreak{}[\allowbreak{}9\allowbreak{}[\allowbreak{}1\allowbreak{}]\allowbreak{}]\allowbreak{}9\allowbreak{}[\allowbreak{}1\allowbreak{}[\allowbreak{}7\allowbreak{}[\allowbreak{}2\allowbreak{}[\allowbreak{}4\allowbreak{}]\allowbreak{}3\allowbreak{}8\allowbreak{}]\allowbreak{}8\allowbreak{}[\allowbreak{}6\allowbreak{}[\allowbreak{}3\allowbreak{}[\allowbreak{}2\allowbreak{}[\allowbreak{}5\allowbreak{}[\allowbreak{}7\allowbreak{}]\allowbreak{}7\allowbreak{}[\allowbreak{}9\allowbreak{}]\allowbreak{}]\allowbreak{}4\allowbreak{}[\allowbreak{}2\allowbreak{}[\allowbreak{}6\allowbreak{}7\allowbreak{}]\allowbreak{}]\allowbreak{}7\allowbreak{}]\allowbreak{}]\allowbreak{}7\allowbreak{}[\allowbreak{}6\allowbreak{}]\allowbreak{}]\allowbreak{}]\allowbreak{}]\allowbreak{}]\allowbreak{}]\allowbreak{}3\allowbreak{}5\allowbreak{}[\allowbreak{}2\allowbreak{}]\allowbreak{}]\allowbreak{}]\allowbreak{}5\allowbreak{}6\allowbreak{}]\allowbreak{}]\allowbreak{}8\allowbreak{}[\allowbreak{}4\allowbreak{}[\allowbreak{}1\allowbreak{}[\allowbreak{}7\allowbreak{}[\allowbreak{}6\allowbreak{}[\allowbreak{}9\allowbreak{}[\allowbreak{}3\allowbreak{}8\allowbreak{}]\allowbreak{}]\allowbreak{}]\allowbreak{}]\allowbreak{}2\allowbreak{}[\allowbreak{}5\allowbreak{}[\allowbreak{}1\allowbreak{}[\allowbreak{}6\allowbreak{}]\allowbreak{}3\allowbreak{}[\allowbreak{}8\allowbreak{}]\allowbreak{}6\allowbreak{}7\allowbreak{}[\allowbreak{}6\allowbreak{}]\allowbreak{}9\allowbreak{}]\allowbreak{}7\allowbreak{}[\allowbreak{}1\allowbreak{}[\allowbreak{}5\allowbreak{}6\allowbreak{}]\allowbreak{}5\allowbreak{}[\allowbreak{}1\allowbreak{}[\allowbreak{}6\allowbreak{}[\allowbreak{}9\allowbreak{}]\allowbreak{}9\allowbreak{}[\allowbreak{}6\allowbreak{}[\allowbreak{}3\allowbreak{}[\allowbreak{}4\allowbreak{}[\allowbreak{}2\allowbreak{}[\allowbreak{}8\allowbreak{}9\allowbreak{}[\allowbreak{}5\allowbreak{}[\allowbreak{}6\allowbreak{}[\allowbreak{}3\allowbreak{}]\allowbreak{}7\allowbreak{}[\allowbreak{}8\allowbreak{}[\allowbreak{}1\allowbreak{}[\allowbreak{}2\allowbreak{}[\allowbreak{}3\allowbreak{}[\allowbreak{}5\allowbreak{}7\allowbreak{}]\allowbreak{}]\allowbreak{}]\allowbreak{}]\allowbreak{}]\allowbreak{}8\allowbreak{}[\allowbreak{}6\allowbreak{}]\allowbreak{}]\allowbreak{}8\allowbreak{}]\allowbreak{}]\allowbreak{}]\allowbreak{}]\allowbreak{}4\allowbreak{}[\allowbreak{}8\allowbreak{}[\allowbreak{}3\allowbreak{}]\allowbreak{}]\allowbreak{}8\allowbreak{}[\allowbreak{}2\allowbreak{}]\allowbreak{}]\allowbreak{}]\allowbreak{}]\allowbreak{}3\allowbreak{}]\allowbreak{}]\allowbreak{}]\allowbreak{}6\allowbreak{}[\allowbreak{}1\allowbreak{}[\allowbreak{}2\allowbreak{}[\allowbreak{}7\allowbreak{}[\allowbreak{}3\allowbreak{}[\allowbreak{}4\allowbreak{}[\allowbreak{}9\allowbreak{}[\allowbreak{}1\allowbreak{}[\allowbreak{}8\allowbreak{}]\allowbreak{}5\allowbreak{}[\allowbreak{}8\allowbreak{}]\allowbreak{}8\allowbreak{}]\allowbreak{}]\allowbreak{}]\allowbreak{}]\allowbreak{}9\allowbreak{}[\allowbreak{}8\allowbreak{}]\allowbreak{}]\allowbreak{}]\allowbreak{}5\allowbreak{}[\allowbreak{}1\allowbreak{}3\allowbreak{}]\allowbreak{}7\allowbreak{}]\allowbreak{}7\allowbreak{}[\allowbreak{}1\allowbreak{}2\allowbreak{}5\allowbreak{}]\allowbreak{}]\allowbreak{}6\allowbreak{}[\allowbreak{}4\allowbreak{}]\allowbreak{}7\allowbreak{}]\allowbreak{}]\allowbreak{}]\allowbreak{}8\allowbreak{}[\allowbreak{}2\allowbreak{}3\allowbreak{}7\allowbreak{}]\allowbreak{}9\allowbreak{}[\allowbreak{}3\allowbreak{}]\allowbreak{}]\allowbreak{}]\allowbreak{}3\allowbreak{}[\allowbreak{}1\allowbreak{}[\allowbreak{}8\allowbreak{}[\allowbreak{}5\allowbreak{}[\allowbreak{}2\allowbreak{}9\allowbreak{}]\allowbreak{}]\allowbreak{}]\allowbreak{}5\allowbreak{}[\allowbreak{}1\allowbreak{}]\allowbreak{}8\allowbreak{}[\allowbreak{}1\allowbreak{}5\allowbreak{}[\allowbreak{}1\allowbreak{}[\allowbreak{}7\allowbreak{}]\allowbreak{}]\allowbreak{}]\allowbreak{}]\allowbreak{}8\allowbreak{}]\allowbreak{}]\allowbreak{}]\allowbreak{}]\allowbreak{}]\allowbreak{}7\allowbreak{}[\allowbreak{}2\allowbreak{}[\allowbreak{}1\allowbreak{}9\allowbreak{}]\allowbreak{}]\allowbreak{}]\allowbreak{}]\allowbreak{}7\allowbreak{}[\allowbreak{}2\allowbreak{}]\allowbreak{}9\allowbreak{}[\allowbreak{}5\allowbreak{}]\allowbreak{}]\allowbreak{}7\allowbreak{}[\allowbreak{}6\allowbreak{}8\allowbreak{}[\allowbreak{}2\allowbreak{}[\allowbreak{}6\allowbreak{}]\allowbreak{}5\allowbreak{}[\allowbreak{}9\allowbreak{}]\allowbreak{}6\allowbreak{}[\allowbreak{}2\allowbreak{}[\allowbreak{}9\allowbreak{}[\allowbreak{}3\allowbreak{}]\allowbreak{}]\allowbreak{}9\allowbreak{}[\allowbreak{}5\allowbreak{}[\allowbreak{}4\allowbreak{}[\allowbreak{}2\allowbreak{}]\allowbreak{}]\allowbreak{}]\allowbreak{}]\allowbreak{}]\allowbreak{}]\allowbreak{}8\allowbreak{}]\allowbreak{}8\allowbreak{}]\allowbreak{}6\allowbreak{}[\allowbreak{}3\allowbreak{}[\allowbreak{}4\allowbreak{}5\allowbreak{}]\allowbreak{}]\allowbreak{}]\allowbreak{}]\allowbreak{}3\allowbreak{}[\allowbreak{}1\allowbreak{}2\allowbreak{}[\allowbreak{}1\allowbreak{}]\allowbreak{}]\allowbreak{}}}\par}\smallskip

\end{document}